\documentclass[pdflatex,sn-mathphys-num]{sn-jnl}

\usepackage{graphicx}%
\usepackage{multirow}%
\usepackage{amsmath,amssymb,amsfonts}%
\usepackage{amsthm}%
\usepackage{mathrsfs}%
\usepackage[title]{appendix}%
\usepackage{textcomp}%
\usepackage{manyfoot}%
\usepackage{booktabs}%
\usepackage{algorithm}%
\usepackage{algorithmicx}%
\usepackage{algpseudocode}%
\usepackage{listings}%
\usepackage{tabularx,booktabs}

\theoremstyle{thmstyleone}%
\theoremstyle{thmstyletwo}%

\theoremstyle{thmstylethree}%

\begin{document}
\addtocontents{toc}{\protect\setcounter{tocdepth}{-5}}

\title[Large cities lose their growth advantage as countries urbanize]{Large cities lose their growth advantage as countries urbanize}

\author[1,2]{\fnm{Andrea} \sur{Musso}}\email{andrea.musso@gess.ethz.ch}

\author[2,3]{\fnm{Diego} \sur{Rybski}}\email{ca-dr@rybski.de}

\author[1,2]{\fnm{Dirk} \sur{Helbing}}\email{dirk.helbing@gess.ethz.ch}

\author[2,4]{\fnm{Frank} \sur{Neffke}}\email{neffke@csh.ac.at}

\affil[1]{\orgdiv{Computational Social Science}, \orgname{ETH Zurich}, \postcode{8092} \orgaddress{\city{Zurich}, \country{Switzerland}}}
\affil[2]{\orgdiv{Complexity Science Hub Vienna}, \postcode{1030} \orgaddress{\city{Vienna},  \country{Austria}}}
\affil[3]{\orgdiv{Leibniz Institute of Ecological Urban and Regional Development}, 
\postcode{01217} \orgaddress{\city{Dresden},  \country{Germany}}}
\affil[4]{\orgdiv{Transforming Economies Lab}, \orgname{Interdisciplinary Transformation University Austria (IT:U)}, \postcode{4040} \orgaddress{\city{Linz},  \country{Austria}}}

\abstract{The share of the world population living in cities with more than one million people rose from 11\% in 1975 to 24\% in 2025 (our estimates). Will this trend towards greater concentration in large cities continue or level off? We introduce two new city population datasets that use consistent city definitions across countries and over time. The first covers the world between 1975 and 2025, using satellite imagery. The second covers the U.S. between 1850 and 2020, using census microdata. We find that urban growth follows a characteristic life cycle. In the early stages of a country's urbanization process, large cities grow faster than smaller ones. At later stages, growth rates equalize across sizes. We use this life cycle to project future population concentration in large cities. Our projections suggest that 38\% of the world population will be living in cities with more than one million people by 2100. This estimate is higher than the 33\% implied by the well-known theory of proportional growth, but lower than the 42\% obtained by extrapolating current trends.}

\maketitle

\newpage
\section{Introduction}
Over the past 50 years, the world population has become increasingly concentrated in large cities. Whether this concentration will continue depends on the relationship between a city's size and its growth rate---a topic on which the literature is divided.

Theories of increasing returns, such as the New Economic Geography \cite{krugman1991increasing, fujita2001spatial} and the evolutionary approach of Pumain \textit{et al.} \cite{pumain2006evolutionary, paulus2004coevolution, pumain2012theorie}, argue that large cities tend to grow faster than small ones. Large cities are typically more innovative and productive \cite{bettencourt2007growtha, bettencourt2014professionala}, have more educated workforces \cite{combes2008spatial, keuschnigg2019urban}, sit more centrally in trade and knowledge networks \cite{pumain2012theorie}, and host more advanced economic activities \cite{balland2020complexa, gomez-lievano2016explainingb}. These factors can plausibly confer them a growth advantage. Based on our novel projection method, even a mild growth advantage---where a 10-fold increase in size corresponds to a 0.7\% increase in yearly growth---would result in $47\%$ of the world population living in 1M$+$ cities by the end of the century. 

Theories of propotional growth argue instead that cities of all sizes tend to grow at similar rates (Gibrat's Law) \cite{sutton1997gibrats, eeckhout2004gibrats, eaton1997cities, gonzalez-val2014newa, henderson2007urbanizationa}. After all, scale also brings costs such as increased congestion \cite{louf2014how, barthelemy2016structure}, crime \cite{oliveira2017scaling, bettencourt2010urban}, disease risk  \cite{rocha2015nonlineara, bilal2021scaling}, and housing expenses \cite{sarkar2019urban, duranton2004chaptera}, and these costs may offset advantages. If city size exhibits no growth advantage, our projections suggest that $33\%$ of the world population will live in 1M$+$ cities in 2100. 

This 14 percentage point gap between the two predictions represents a difference of 1.4 billion people and thus has significant implications for future planning and development. In this paper, we analyze which of these two scenarios is more likely, using two newly built city population datasets. The first dataset, built from satellite-derived grids \cite{schavina2023ghsl}, spans 1975-2025 and covers 99 countries, together representing 94\% of the world population. The second, built from census microdata \cite{ruggles2024ipums, manson2024ipums, berkes2023censusa}, spans 1850-2020 and covers the United States (USA) urban system over nearly its entire history. 

These datasets substantially extend the spatial and temporal coverage of previous efforts to build such harmonized data \cite{pumain2015multilevel}. Before the advent of satellite imagery and geospatial processing tools, assembling harmonized datasets of city sizes across countries and years was extremely difficult, primarily because national data collection efforts are not designed to produce outputs that are comparable across countries or consistent over time. As a result, most empirical studies of urban growth focused on just a handful of countries over limited time periods. The conclusions of these studies are varied, some supporting proportional growth \cite{eeckhout2004gibrats, eaton1997cities, gonzalez-val2014newa, henderson2007urbanizationa} and others not \cite{guerin-pace1995ranksizea, desmet2017settlementa, michaels2012urbanizationa, rozenfeld2008lawsa, black2003urban, soo2007zipfs}. Integrating these results has proved difficult due to substantial differences in methodology \cite{cottineau2017metazipf}, leaving us with an incomplete understanding of how urban growth evolves over time \cite{verbavatz2020growtha}.

Using these new datasets, we sharpen our understanding of temporal trends in urban growth. First, we show that proportional growth and increasing returns are better understood as two phases of the same underlying process, not competing realities. In the early phase of urbanization, large cities enjoy a strong growth advantage, consistent with increasing returns. As an urban system matures, this advantage weakens, and the growth rates of large and small cities converge, consistent with proportional growth. Second, we show that this size-growth relationship translates into systematic changes in the shape of a country's city-size distribution over time \cite{zipf1949humana, batty2006rank, rybski2023auerbach}. When large cities have a growth advantage, the distribution stretches and its rank-size slope (a spline-based analogue of the Zipf exponent \cite{saichev2010theory}) increases. Third, combining this mechanism and a novel projection method, we project that by 2100, 38\% of the world's population will live in 1M$+$ cities. This projection lies between the proportional-growth (33\%) and increasing-returns (47\%) benchmarks, and below an extrapolation of current trends (42\%).

\section{Results}

\begin{table}[]
    \centering
\begin{tabular*}{0.8\linewidth}{@{\extracolsep{\fill}}lccccc}
\\[-1.8ex]\hline\hline \\[-1.8ex]Dataset & City-Year Obs. & Countries & Frequency (years) & Time Period \\
\midrule
Global cities & 1,604,593 & 99 & 5 & $1975-2025$ \\
USA cities & 26,902 & 1 & 10 & $1850-2020$ \\
\hline\hline \\[-1.8ex]\end{tabular*}
    \caption{Description of the two large-scale city population datasets created for the analysis. The Global cities dataset covers 99 countries around the world (amounting to 94\% of the world population in 2025) between 1975 and 2025. Its primary source is the GHSL 2023 data package \cite{schavina2023ghsl}. The USA cities dataset covers the continental USA between 1850 and 2020. Its primary sources are IPUMS USA \cite{ruggles2024ipums}, IPUMS NHGIS \cite{manson2024ipums}, and the Census Place Project \cite{berkes2023censusa}.}
    \label{tab:1}
\end{table}

Our city population datasets (Table \ref{tab:1}) define cities geographically, as clusters of contiguous built-up areas or regions of high population density \cite{makse1995modelling, rozenfeld2008lawsa, arcaute2015constructing} (Methods \ref{methods:constructing_cities}). This definition has three main advantages: (i) it dynamically adjusts as urban areas expand; (ii) it is consistent across space and time, facilitating robust comparative analyses; and (iii) it is unaffected by political boundary changes such as municipal mergers. The trade-off is that this definition captures morphological clusters rather than functional urban areas, so satellite settlements tied to large cities by commuting flows may appear as separate clusters (see SI 5 for further discussion).

\begin{figure}[ht!]
    \centering
    \includegraphics[width=0.95\textwidth]{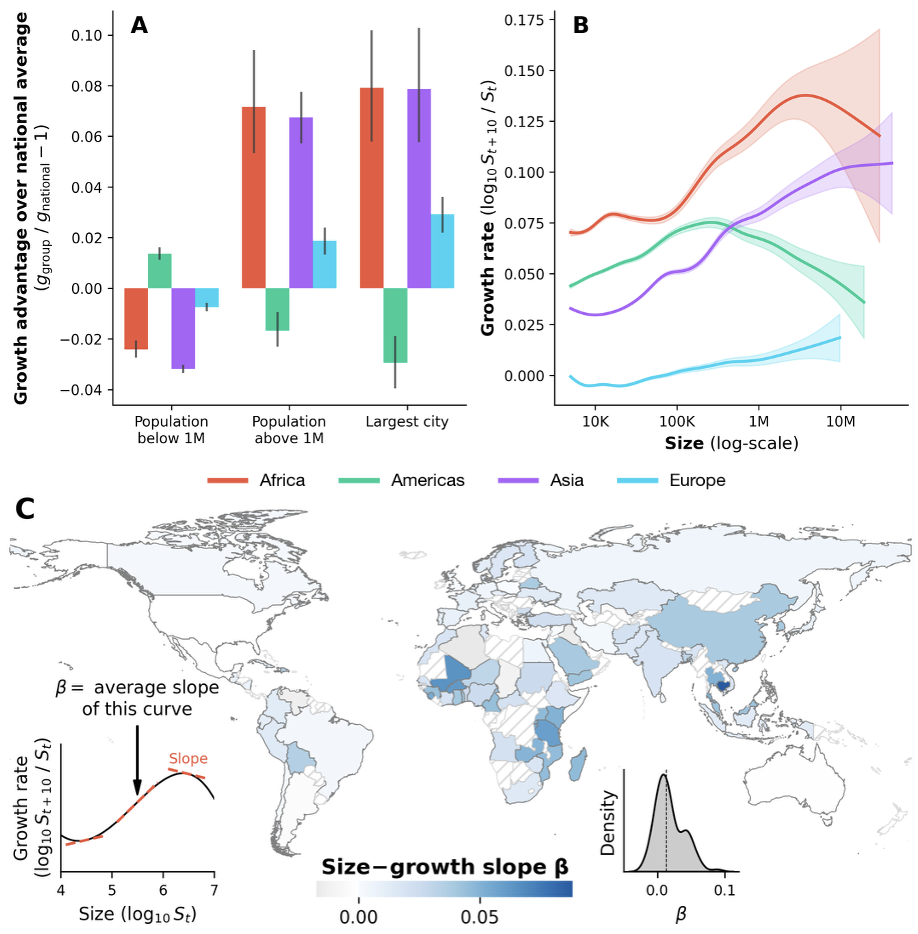}
    \caption{The growth advantage of large cities varies systematically across regions: it is strong in Asia and Africa and weak in Europe and the Americas. \textbf{(A)} We compare the average growth rate of a country's cities, denoted by $g_{\text{national}}$, with the average growth rate of specific sub-groups of its cities (e.g., the largest city, 1M$+$ cities, etc.), denoted by $g_{\text{group}}$. Each bar in the chart shows the ratio $g_{\text{group}} \ / \ g_{\text{national}} - 1$ for a given region. \textbf{(B)} \emph{Size-growth curve} by region. These curves are obtained by fitting a penalized cubic B-spline ($\lambda = 100$) to the relationship between city log-size in year $t$ and city log-growth between year $t$ and $t + 10$ (Methods \ref{methods:measurement}). \textbf{(C)} \emph{Size-growth slopes} $\beta$ by country. $\beta$ is obtained by first estimating the national size-growth curve (as in panel (B)), and then averaging the local slope of this curve across the size spectrum (see left inset and Methods \ref{methods:measurement}). The map shows the mean value of $\beta$ between 1975 and 2025; hatched indicates no data. The right inset shows a kernel density estimate of the distribution of $\beta$ across countries (dotted line = median).}
    \label{fig:1}
\end{figure}

Our analysis of these datasets reveals substantial global variation in urban growth patterns. Between 1975 and 2025, an average 1M$+$ city in Asia/Africa outgrew the national average by $\sim 7\%$ (Fig. \ref{fig:1}A). In contrast, Europe's large cities grew modestly faster than the rest (Fig. \ref{fig:1}A-B), and in the Americas, city growth displayed an inverted-U-shaped trend, with 1M$+$ cities growing 1.6\% slower than their national average (Fig. \ref{fig:1}A-B). 

These patterns are also visible at the country level. Figure \ref{fig:1}C maps national size-growth slopes $\beta$ (definition in the caption of Figure \ref{fig:1}C). A positive $\beta$ indicates that growth increases with size or, in other words, that large cities have a growth advantage. This growth advantage varies significantly across countries, with $\beta$ ranging from $-0.02$ to $0.1$ (median $\approx 0.012$; Fig. \ref{fig:1}C inset). Notably, $\beta$s are generally higher in Asia and Africa and lower in Europe and the Americas.

\begin{figure}[ht!]
    \centering
    \includegraphics[width=0.95\textwidth]{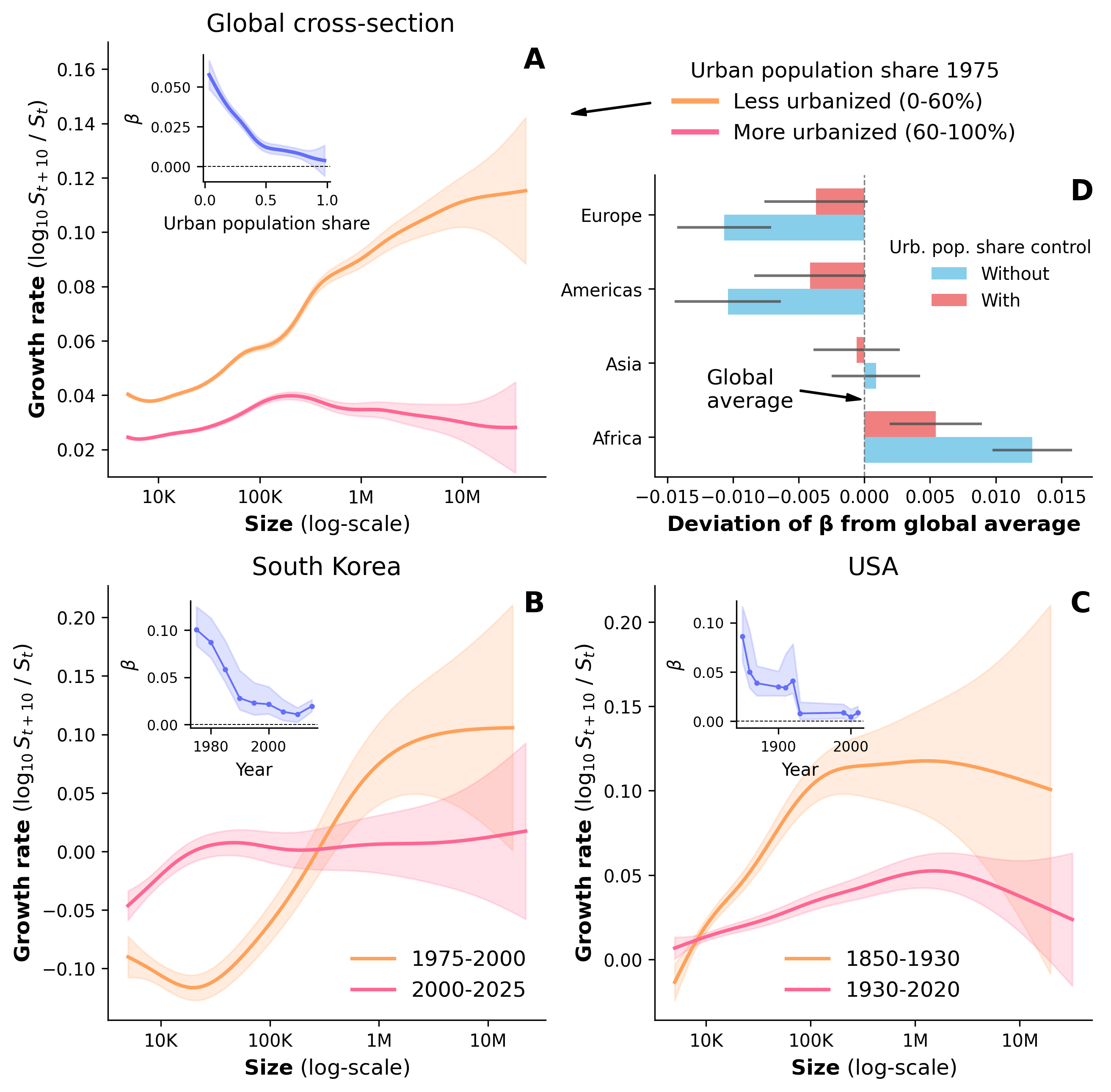}
    \caption{The growth advantage of large cities weakens as countries urbanize. \textbf{(A)} Size-growth curve by urbanization group (Methods \ref{methods:measurement}). (Inset) $\beta$ vs.\ urban population share (fit with a penalized cubic B-spline with $\lambda = 100$). \textbf{(B-C)} Size-growth curves in South Korea and the USA across different time periods. (Insets) $\beta$ vs.\ time (bootstrapped confidence intervals $n = 1000$). \textbf{(D)} Country-level regression with regional dummies ($\beta_{t, \text{country}} \sim \delta_{\text{region}} + \text{urban population share}_{t, \text{country}}$). Regional differences in $\beta$ become significantly smaller once we control for the level of urbanization.}
    \label{fig:2}
\end{figure}

This regional divide can be understood as part of a universal dynamic in which the growth advantage of large cities weakens as a country urbanizes. This inverse relationship is evident in both cross-sectional and longitudinal data (Fig. \ref{fig:2}).

Across countries, a lower national share of urban population correlates with higher $\beta$ (Fig. \ref{fig:2}A inset and Table \ref{tab:2}). Put differently, growth rates increase rapidly with city size in less urbanized countries, whereas the size-growth relationship is nearly flat in more urbanized ones (Fig. \ref{fig:2}A). The numbers speak clearly: between 1975 and 2025, 1M$+$ cities in more urbanized countries grew at the national average rate, while 1M$+$ cities in less urbanized ones grew 7.3\% faster.

Within countries, $\beta$ declines as urbanization rises. A regression with country fixed effects shows that a 20\% increase in a country's urban population share is associated with a $\sim 0.01$ reduction in $\beta$ (Table \ref{tab:2}). This pattern is clearly observed in South Korea and the USA, two countries for which our data cover a large window of the urbanization process. Both countries saw large cities grow substantially faster than smaller ones in the early phase of urbanization, but as urbanization progressed, this advantage largely vanished (Fig. \ref{fig:2}B-C).

These results help explain the regional differences in $\beta$ observed in Figure \ref{fig:1}. A simple regression shows that, when controlling for urbanization level, regional dummies converge toward the global average, with differences becoming either statistically insignificant or much smaller (Fig. \ref{fig:2}D). This indicates that regional differences in $\beta$ largely reflect each country's stage in the urbanization process.

\begin{figure}[ht!]
    \centering
    \includegraphics[width=0.95\textwidth]{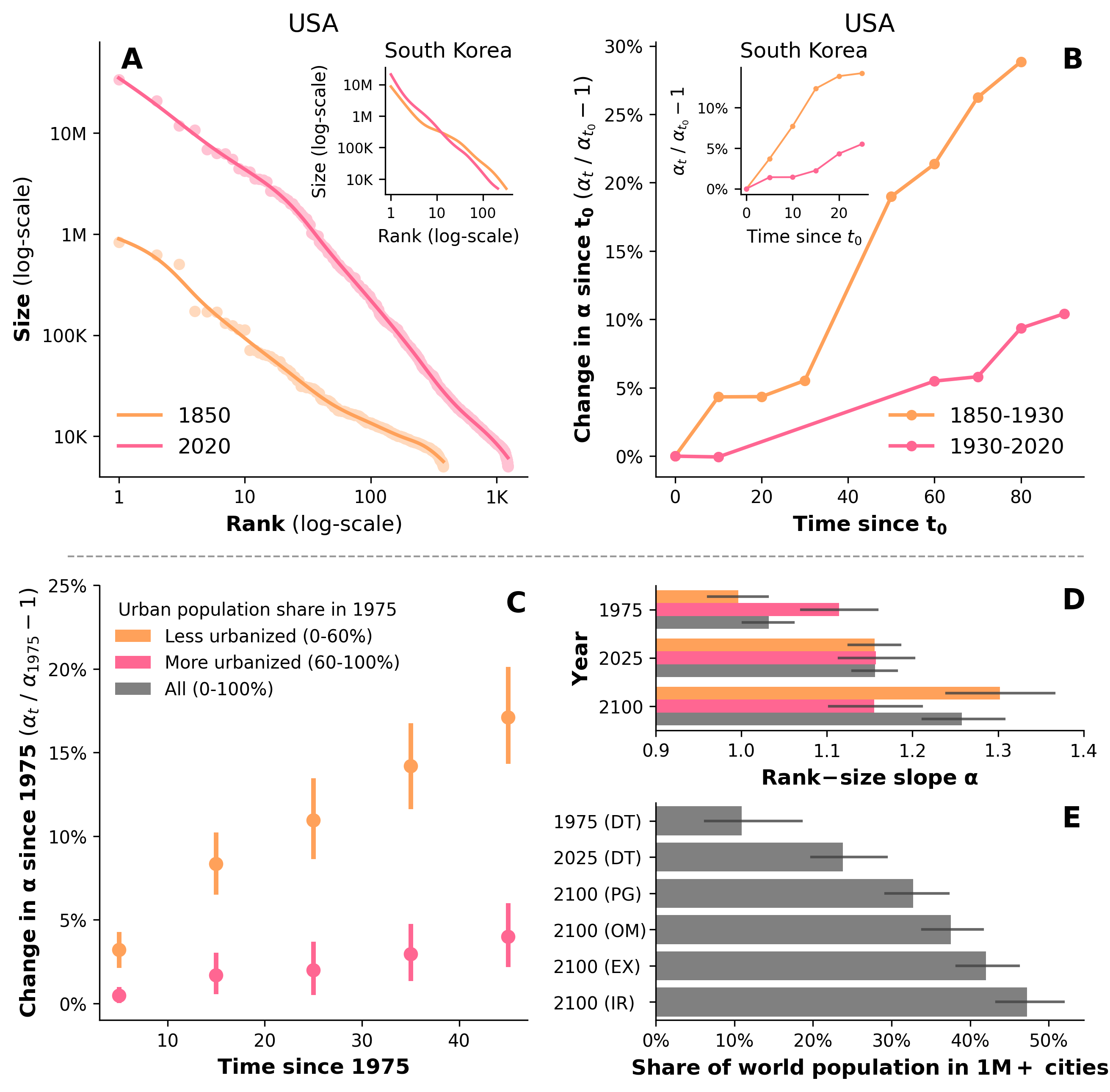}
    \caption{Historical trends and projections for national city size distributions. \textbf{(A)} Rank-size curves for the USA and South Korea. These curves are obtained by fitting penalized cubic B-splines ($\lambda = 1$) to the relationship between city log-rank and city log-size in a fixed year. The USA plot also displays a scatter of the underlying data points. \textbf{(B-C)} The rank-size slope $\alpha$ is estimated by averaging the local slope of the rank-size curve across ranks and then taking the absolute value (Methods \ref{methods:measurement}). Note that $\alpha$ is a spline-based analogue of the Zipf exponent. \textbf{(B)} Change in $\alpha$ for South Korea and the USA from a base year $t_0$ (South Korea $t_0 = 1975,2000$; USA $t_0=1850,1930$). \textbf{(C)} Change in average $\alpha$ across more/less urbanized countries since 1975 (binned scatter plot by year with bootstrapped confidence intervals; $n = 1000)$. \textbf{(D)} Average $\alpha$ across groups of countries in different years. 2100 values are projected using the technique in Methods \ref{methods:projections}. \textbf{(E)} Share of the world population in 1M$+$ cities in 1975, 2025, and 2100. 2100 values are projected under four urban growth models, which differ in their estimates for the future trajectories of $\beta$: proportional growth (PG; $\beta = 0$), increasing returns (IR; $\beta = 0.03$), current trend extrapolation (EX; $\beta$ equal to the country's average between 1975-2025), and our model (OM; time-varying $\beta$ according to the country's urbanization). See Methods \ref{methods:projections}.} 
    \label{fig:3}
\end{figure}

The size-growth patterns observed in Figure \ref{fig:2} change national city size distributions.
We can quantify this change by looking at the distribution's rank-size curve (Fig. \ref{fig:3}A), or more precisely at the absolute value of its slope, $\alpha$ (exact definition of $\alpha$ in the caption of Figure \ref{fig:3}B-C). If $\beta$ is positive, large cities have a growth advantage, the left tail of the rank-size curve rises faster than the right one, the curve steepens, and $\alpha$ increases. Thus, if $\beta$ declines with urbanization (as in Fig. \ref{fig:2}), $\alpha$ should increase at a decelerating rate. Our data confirm this: $\alpha$ increased in both the USA and South Korea (Fig. \ref{fig:3}B), and it did so faster early on. Further, since 1975, $\alpha$ grew $18\%$ on average across less urbanized countries, against $4\%$ across more urbanized ones (Fig. \ref{fig:3}C). 

With the mechanism in hand---$\beta$ shapes $\alpha$, and $\beta$ weakens as countries urbanize---we move from description to prediction by projecting trajectories of national $\alpha$s to 2100 (Methods \ref{methods:projections}). Our projections suggest that growth in $\alpha$ will slow over the coming decades. Between 1975 and 2025, $\alpha$ grew by 2.3\% per decade on average across the countries in our sample. Between 2025 and 2100 that same average is projected to grow at 0.9\% per decade. Further, $\alpha$ will continue to grow faster in countries that were less urbanized in 1975. By 2100, their average $\alpha$ is projected to be $16\%$ larger than that of the more urbanized group, while it was $12\%$ smaller in 1975 (Fig. \ref{fig:3}D). 

Translating projected $\alpha$s into population shares, we estimate that 38\% of the world population will live in 1M$+$ cities by 2100 (Fig. \ref{fig:3}E; Methods \ref{methods:projections}). This projection sits below an extrapolation of current trends ($42\%$; Fig. \ref{fig:3}E) and above the proportional-growth benchmark (33\%; Fig. \ref{fig:3}E), consistent with a weakening growth advantage of large cities as urbanization progresses. 

\section{Discussion}

Using a robust geographic definition of cities and a comprehensive database spanning several countries and historical periods, we show that urban growth follows a typical life cycle. Early in urbanization, large cities experience a strong growth advantage, stretching the city size distribution. As urbanization progresses, this growth advantage weakens and the distribution stabilizes. We use this model to forecast urban concentration at the end of the century. Relative to an extrapolation of 1975-2025 trends, our model projects 450 million fewer residents in 1M$+$ cities by 2100 ($-4.4\%$); relative to proportional growth, 490 million more ($+4.8\%$).

Our study has some limitations. First, our city definition is geographic rather than functional. It tracks clusters of built-up area or high population density, making it well suited for comparative analyses across countries and years. However, because it separates central cores from their surrounding suburbs, it may fail to fully capture the true gravitational pull of large cities. As a result, this definition might underestimate the growth advantage of large functional urban areas, especially in urbanized countries where advanced transport networks drive growth from city cores into surrounding suburbs and satellite settlements. Our supplementary analysis suggest that this is the case: size-growth slopes estimated using geographic urban areas are typically flatter than those estimated using functional urban areas (SI 5.1). However, this underestimation is small relative to the overall trends, such that our findings do not substantially change when shifting the analysis to the level of functional urban areas. Accordingly, our projections should be read as conservative estimates of future concentration in functional urban areas.

Second, we use national borders to define urban systems. This definition is imperfect, especially in settings with strong cross-border integration. Our choice is dictated by both convenience and substance. With respect to convenience, countries are the territorial units that are most readily comparable in global analyses and that have the most widely available longitudinal information on their urban systems, both historically and for projections into the future. In more substantive terms, national borders strongly constrain the migration patterns, trade, institutions, and infrastructure that profoundly shape urban development \cite{ades1995tradea}. Developing a more principled approach to defining urban systems, perhaps through migration networks \cite{bettencourt2020demographyb}, is nevertheless a promising area for future research. To partially accommodate the limitations of relying on national borders to delineate urban systems, we provide a supplementary analyses at the level of UN M49 sub-regions (SI 4.1). Its findings align with our country-level ones, corroborating the declining growth advantage of large cities also at intermediate spatial scales.

Third, we focus on population, neglecting other aspects of city size, such as GDP, innovation, or employment. Population is the natural starting point for our analysis because it is the main measure of city size, it has been used for centuries, it is broadly comparable across countries, and it is central to the Gibrat, Zipf, and scaling literatures \cite{batty2006rank, bettencourt2007growtha, gabaix1999zipfs, pumain2015multilevel, cottineau2017metazipf}. However, extending our work to economic outcomes and measuring how scaling laws change over time is a valuable direction for future research \cite{pumain2006evolutionary, bettencourt2020urban, balland2020complexa}. The IPUMS full-count census data used in this paper are an excellent starting point for such work \cite{ruggles2024ipums, berkes2023censusa}.

Fourth, our framework is phenomenological rather than causal. We document a robust empirical regularity and show how it may be used to improve projections. However, we do not isolate the causal mechanisms underpinning this regularity. Several mechanisms could cause the weakening of the growth advantage of large cities, including slowing rural-to-urban migration, reduced imbalances in migration flows \cite{reia2022spatial, verbavatz2020growtha, bettencourt2020demographyb}, rising congestion and housing costs \cite{sarkar2019urban}, diffusion of economic activity \cite{pumain1997city, pumain2012theorie}, and the redistribution of metropolitan growth toward suburban peripheries (suburbanization) \cite{soo2005zipfs}. In SI 5, we provide a brief discussion of the role of suburbanization, showing that its effect is real but insufficient to fully explain this phenomenon. Distinguishing the relative contributions of other causes is a promising area of future work.

Fifth, our projections are conditional on the absence of major external shocks --- the standard \emph{ceteris paribus} assumption. Because our model is not structural, we cannot explicitly integrate external shocks, such as geopolitical disruptions, climate-driven migration, pandemics, or technological revolutions. 
Any of these forces could, in principle, reshape urban trajectories in ways our model does not capture. For example, advances in urban planning and sanitation could reduce the diseconomies of scale that constrain large cities. Climate change could intensify migration toward hubs. That said, the urbanization pattern we document has proven remarkably robust. It has persisted through a profoundly volatile century, withstanding disruptions such as the automobile revolution, the internet, the rise and fall of geopolitical blocs, the construction of modern urban infrastructure, and the globalization of trade.

Despite the above limitations, our results have relevant implications when viewed through the lens of urban scaling theory \cite{rybski2019urban}. Because many urban outcomes scale nonlinearly with population, reallocating people across cities of different sizes changes aggregate outcomes (see SI 2.3 and \cite{bettencourt2007growtha, ribeiro2021association, gomez-lievano2012statistics}). Our results suggest that as urbanization progresses, this reallocation becomes less skewed toward the largest cities and more evenly distributed across the urban system. The policy implications are mixed. On the one hand, if productivity scales super-linearly with city size, a slowdown in the growth of large cities implies a slowdown in reallocation-driven productivity growth \cite{bettencourt2007growtha, keuschnigg2019urban, roca2017learning}. On the other hand, if emissions, urban heat islands, or other environmental burdens rise super-linearly with city size, the same slowdown may ease some environmental pressures \cite{sarzynski2012bigger, yang2025scaling}. Put simply, whether urban concentration or equalization is preferable depends on which of these trade-offs policymakers prioritize---a question that research can inform but not resolve.

\section{Methods}
\subsection{Data}
\label{methods:data}
This paper builds on several datasets, all of which are publicly available. Below we provide a high-level overview of the main datasets. In Methods \ref{methods:code_and_data_availability}, we document how to retrieve them. \\

\noindent \textbf{Population and Urbanization data}: The population projections come from the United Nations World Population Prospect 2024 \cite{undesa2024wpp}, the History database of the Global Environment 2023 \cite{pbl2023hyde}, and Gapminder \cite{gapminder2023systema, gapminder2022population}, with processing from Our World in Data \cite{owid2024population}. We use projections under the medium fertility scenario. The urbanization data come from Chen \textit{et al.} \cite{chen2022updating}. We use their World Bank-based annual projections under SSP2 (``middle of the road''), which extend the World Bank series to 2100.  This data relies on national statistical definitions of ``urban'', which vary across countries and differ from our morphological definition. We adopt these projections for three reasons. First, they are independently validated. Second, they are directly connected to future projections. Third, they provide more reliable estimates of the urban population share than GHSL grids, which are known to underestimate rural populations \cite{schavina2023ghsl, lang-ritter2025global}. We therefore use GHSL grids only to compare urban areas, not to estimate the overall urban population share. Our results are robust to the use of a different definition of urban population share (SI 4.2).
\footnote{} \\

\noindent \textbf{Country borders}: Boundaries as of 2019, from the CShapes database \cite{schvitz2022mapping}. \\

\noindent  \textbf{Global grids}: Our global cities dataset is based on the population (pop) and degree-of-urbanization (smod) grids from the Global Human Settlement Layer (GHSL) 2023 data package \cite{schavina2023ghsl}. These grids estimate population counts and urbanization levels for 1km-by-1km cells covering the whole planet by combining satellite imagery with census data. \\

\noindent \textbf{USA grids}: Our USA cities dataset is based on custom grids derived from census place population estimates from various data sources.
For the 1990-2020 period, we use estimates from the National Historical Geographic Information System (IPUMS NHGIS) \cite{manson2024ipums}. For the 1850-1940 period, we reconstruct census place population estimates using the IPUMS USA full count census data \cite{ruggles2024ipums} with geocoding from the Census Place Project \cite{berkes2023censusa}. To do so, we match over 500 million individual census records to approximately 40,000 census places. We then aggregate these records to estimate the population of each census place.
Because this historical data comes from century-old handwritten documents, it contains numerous inconsistencies, such as census places disappearing and reappearing over time. We correct for these issues using several preprocessing steps that leverage individual migration data, as detailed in SI 1.1.

We use these clean census place population estimates to build population grids suitable for the City Clustering Algorithm. 
For each year $t$, we create a 1km-by-1km grid covering the continental US. Each grid cell $c_j$ receives an initial population $\text{pop}_{\text{init}}(c_j)$, given by the sum of the population of all census places whose geographic center falls within that cell. We then smooth this initial grid using a spatial convolution kernel: the final population of cell $c_i$ becomes a distance-weighted average of the initial population of neighboring cells $c_j$, with weights decreasing exponentially with distance $d_{c_ic_j}$:
\begin{equation}
    \text{pop}(c_i) = \frac{1}{\sum_{j} e^{-\eta \cdot d_{c_ic_j}}} \sum_{j} e^{-\eta \cdot d_{c_ic_j}} \cdot \text{pop}_{\text{init}}(c_j) \ .
\end{equation}
The decay parameter $\eta$, which controls how quickly population decays around a census place, is set to $\eta = 0.2$, similar to estimates in prior work \cite{batistaesilva2020uncovering}. 

\subsection{Constructing cities: the City Clustering Algorithm}
\label{methods:constructing_cities}

\begin{figure}[h!]
    \centering
    \includegraphics[width=\textwidth]{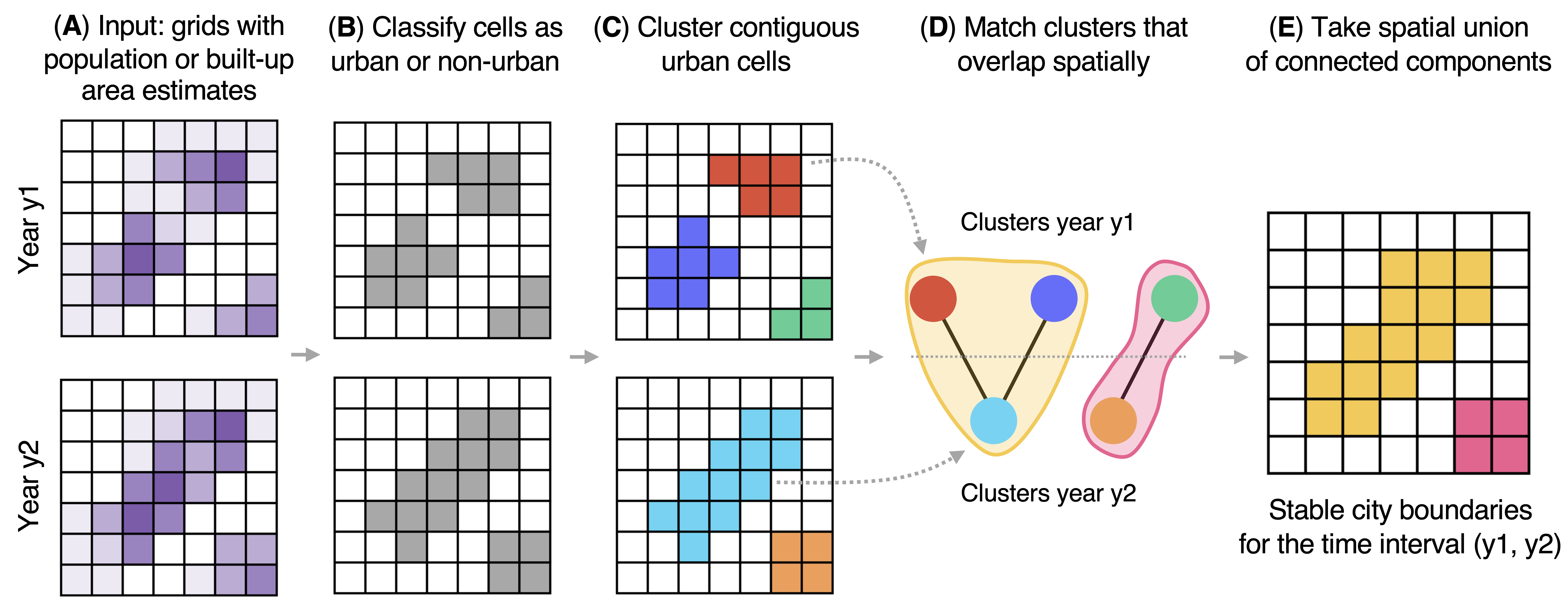}
    \caption{
    The City Clustering Algorithm \cite{rozenfeld2008lawsa} computes stable city boundaries for a time interval $(y_1, y_2)$. This algorithm proceeds in five steps: \textbf{(A)} It takes two grids as input, one for each year, containing estimates of population or built-up area. \textbf{(B)} It classifies grid cells as either urban or non-urban using a threshold on these population/built-up area estimates. \textbf{(C)} It groups contiguous urban grid cells to form clusters. \textbf{(D)} It matches clusters across years when they overlap spatially, forming a bipartite graph. \textbf{(E)} It defines a city's boundary as the spatial union of all clusters within a single connected component of the graph.
    }
    \label{fig:4}
\end{figure}

To construct cities from the USA and global grids, we use the City Clustering Algorithm \cite{rozenfeld2008lawsa}. This algorithm identifies stable city boundaries between two years $t_1$ and $t_2$ based on the city's geographic extent, allowing for robust measurement of population growth and comparable city definitions across countries and times. The algorithm, illustrated in Figure \ref{fig:4}, involves five steps:
\begin{enumerate}
    \item \textbf{Classify urban cells}: for each year independently, we classify all grid cells as either ``urban'' or ``non-urban'' using population density or degree-of-urbanization thresholds.
    \item \textbf{Form initial clusters}: for each year independently, we group contiguous \emph{urban} cells into initial clusters using a flood fill algorithm \cite{hoshen1976percolation}.
    \item \textbf{Match clusters over time}: for a pair of years $t_1 \leq t_2$, we construct a bipartite graph $G = (A,B)$, where nodes in part $A$ are clusters in year $t_1$, and nodes in part $B$ are clusters in year $t_2$. We connect clusters with an edge if and only if they overlap spatially. 
    \item \textbf{Define stable city boundaries}: we extract the connected components of the bipartite graph $G$ that have at least one cluster in year $t_1$ \footnote{We do so to avoid ``new cluster bias'' (see below)}. Each component represents a single, evolving city that existed throughout the $(t_1, t_2)$ period. The city's boundary for the period is the spatial union of all clusters within the component (from both $t_1$ and $t_2$).
    \item \textbf{Calculate population growth}:  for each city, we calculate its population in years $t_1$ and $t_2$ by summing the population of all grid cells that fall within its boundary. The ratio of these two population values gives us the city's growth rate over the period.
\end{enumerate}

While the above algorithm largely follows the same approach as the original paper \cite{rozenfeld2008lawsa}, we introduce some minor improvements and clarifications to its implementation.

First, using a bipartite graph to match clusters (Step 3) provides a more systematic and scalable approach than the manual specifications suggested in the original paper. It improves conceptual clarity, simplifies implementation, and naturally handles events like the merger or split of more than two clusters.

Second, our implementation is less ambiguous with respect to the ``cell reclassification problem''. This problem occurs when cells flip their classification, moving from ``non-urban'' in year $t_1$ to ``urban'' in year $t_2$ (or vice versa). These flips do not require any dramatic transformation. As we use thresholds to determine ``urban'' and ``non-urban'' status, even small changes in population or built-up area can cause a flip.  But if not treated carefully, flips may lead to biased growth estimates. For example, imagine a city where hundreds of grid cells flip from ``non-urban'' to ``urban''. If we simply count the population in these newly ``urban'' cells as if they were entirely new additions to the city, we grossly overestimate its growth. Our solution (Steps 4 and 5) is to define a stable city boundary for the entire period and measure population changes only within that boundary. This isolates true population growth from the artifacts of reclassification. While this solution was adopted in the original CCA paper, the implementation is framed ambiguously and the problem is not mentioned. 

Third, our implementation addresses a form of selection bias that we call ``new cluster bias''. This bias arises if one includes urban clusters that appear in year $t_2$ but have no predecessors in year $t_1$. These ``new'' clusters represent only the fastest-growing locations among a larger pool of similar areas, many of which did not grow enough to become urban clusters. Including these successful outliers while ignoring the rest would upwardly bias the estimated growth rates for small cities. Our method avoids this by analyzing only components that contain at least one cluster from the start of the period (Step 4).

To produce the final datasets for our analysis, we apply the CCA to the USA and global grids with specific hyper-parameters. We classified urban cells using distinct criteria for the USA and the global grids. For the USA grids, we use a simple population threshold, classifying a cell as ``urban'' if it has more than 50 people. For the global grids, we use degree-of-urbanization estimates from the smod grid, classifying a cell as urban if it belongs to a ``semi-dense urban cluster'' or higher (cell value $\geq 22$). Further, we applied a common city threshold to both datasets, filtering out clusters with fewer than 5000 inhabitants. Finally, we excluded from the global cities dataset: (i) Nepal and Myanmar due to data quality issues; (ii) countries with fewer than 50 cities; (iii) countries not present in the urbanization data (\cite{chen2022updating}). In SI 3.1 and 3.2, we discuss these hyperparameter choices in more detail and conduct sensitivity analyses confirming that our results are robust to reasonable variations in their values.

\subsection{Analysis}
\subsubsection{Analytical framework}
\label{methods:theory}
We use a simple analytical framework to guide our analysis. While this does not constitute a causal theory of urban development, it provides a rigorous mathematical formalization that links size-growth slopes, rank-size slopes, and the population share in million-plus cities.This analytical framework rests on two simplifying assumptions. First, a city's size $S(t)$ at time $t$ is a power-law function of its (descending) rank $R(t)$, meaning that the rank-size curve ($x=$log-rank;$y=$log-size) is a straight line with slope $a_t$\footnote{Note that with this definition the complementary cumulative density function (CCDF) of the city size distribution is $P(S > x) \propto x^{-1/a_t}$. Other authors define the exponent with its reciprocal $\zeta_t = 1/a_t$ so that the CCDF is $P(S > x) \propto x^{-\zeta_t}$.}:
\begin{equation}
    S(t) \propto R(t)^{-a_t}
\end{equation}
Second, a city's growth rate between $t$ and $t + 10$ is a power-law function of its size $S(t)$, meaning that the size-growth curve ($x=$log-size; $y=$log-growth) is a straight line with slope $b_t$:
\begin{equation}
g(t, S(t)) = \frac{S(t + 10)}{S(t)} \propto S(t)^{b_t} \ .
\end{equation}
Under these assumptions, $a_t$ and $b_t$ are related by a simple equation:
\begin{equation}
\label{eq:power_law_change}
a_{t + 10} = a_t \cdot (1 + b_t) \ .
\end{equation}
For $t_1 > t_0$, this equation generalizes to:
\begin{equation}
\label{eq:power_law_change_iterated}
a_{t_1} = \exp\Big(\log(a_{t_0}) + \sum_{s \in t_0, t_0 + 10, \cdots, t_1 - 10} \log(1 + b_s)\Big) \ .
\end{equation}
Furthermore, $a_t$ is linked to the share of the \emph{urban} population living in 1M$+$ cities, $m_t$. In fact, the probability of observing a city larger than size $x$ in year $t$ is given by $P(S(t) > x) \propto x^{-1/a_t}$, so:
\begin{align}
\label{eq:shares_vs_rank_size_slope}
m_t &= \int_{z}^{x_{t, \max}} P(S(t) > x) dx  \ / \  \int_{x_{t, \min}}^{x_{t, \max}} P(S(t) > x) dx \\
&=\frac{x_{t, \max}^{1 - 1/a_t} - z^{1 - 1/a_t}}{x_{t, \max}^{1 -1/a_t} - x_{t, \min}^{1 - 1/a_t}} \ .
\end{align}
Here, $x_{t, \max}$  and $x_{t, \min}$ are upper and lower bounds on the size of a country's cities, and $z =10^6 = \text{1M}$.

In sum, this analytical framework provides simple closed-form equations relating the growth advantage of large cities $(b)$ to the concentration of population within them ($a$, $m$).

\subsubsection{Empirical measurement}
\label{methods:measurement}
The above framework highlights $a$ and $b$ as the core parameters governing the evolution of urban systems. We estimate these parameters as follows:
\begin{itemize}
    \item[($a$)] The rank-size slope $\alpha$ is our empirical estimate for the parameter $a$. To obtain $\alpha$ we first estimate the rank-size curve $h$ by fitting a penalized cubic B-spline (with penalty $\lambda = 1$) to city log-rank vs.\ city log-size data points:
    \begin{equation}
    \log_{10}\big(S(t)\big) = h\big(\log_{10}(R(t))\big) \ .
    \end{equation}
     Then we take the absolute value of the mean derivative of $h$ over the log-rank spectrum:
     \begin{equation}
    \alpha =  \frac{1}{r_{\max} - r_{\min}} \Big| \int_{r_{\min}}^{r_{\max}} h'(r)dr \Big| = \Big|\frac{h(r_{\max}) - h(r_{\min})}{r_{\max} - r_{\min}}\Big| \ , \ r = \log_{10}(R) \ .
    \end{equation}
    \item[($b$)] The size-growth slope $\beta$ is our empirical estimate for the parameter $b$. As above, we obtain $\beta$ by first estimating the size-growth curve $f$, a penalized cubic B-spline (with penalty $\lambda = 100$) fit to the city log-size vs.\ city log-growth data points:
    \begin{equation}
    \log_{10}\big(g(t, S(t))\big) = f\big(\log_{10}(S(t))\big) \ .
    \end{equation}
    Then we take the mean derivative of $f$ over the observed log-size spectrum:
\begin{equation}
    \beta = \frac{1}{s_{\max} - s_{\min}}\int_{s_{\min}}^{s_{\max}} f'(s)ds = \frac{f(s_{\max}) - f(s_{\min})}{s_{\max} - s_{\min}} \ , \ s = \log_{10}(S) \ .
\end{equation}
\end{itemize}

We also estimate size-growth curves for groups of countries (such as Europe or less urbanized countries in 1975). The procedure to estimate a group-level size-growth curve $f_g$ comprises three steps. First, we normalize each city's growth rate $g(t, S_{ic}(t)) = S_{ic}(t+10) / S_{ic}(t)$ by the average city growth rate of its country $c$:
\begin{equation}
g_{c}(t) = \sum_{i \in c}S_{ic}(t + 10) / \sum_{i \in c}S_{ic}(t) = \sum_{i} \frac{S_{ic}(t)}{\sum_{i \in c}S_{ic}(t)}\cdot g(t, S_{ic}(t)) \ .
\end{equation}
Second, we pool the normalized growth rates from all countries in the group and fit a single penalized cubic B-spline $\bar{f}_g$ (with $\lambda = 100$):
\begin{equation}
\log_{10}\Big(\frac{g(t, S_{ic}(t))}{g_c(t)}\Big) = \bar{f}_g\big(\log_{10}(S_{ic}(t))\big) \ .
\end{equation}
Third, we recenter the function $\bar{f}_g$ by the group's average growth rate:
\begin{equation}
f_g = \bar{f}_g + \frac{1}{|g|}\sum_{c \in g}\log_{10}(g_{c}(t)) \ .
\end{equation}

Other authors use ordinary least squares (OLS) regression to estimate the parameters $a$ and $b$. Our approach has two advantages over the OLS approach. First, the resulting parameter estimates are more robust to arbitrary design choices, such as the lower threshold for city size. Second, the estimates align more closely with the equations derived from the analytical framework, even when the data deviate from the framework's ideal assumptions (SI Figs. S4 and S5). These advantages can be explained by how each approach calculates average slopes. Our approach calculates the average slope of a given curve by putting equal weights on each part of the $x$-range. The OLS approach, in contrast, weights parts of the $x$-range by the density of data that are within them. Because small cities dominate city population data, the OLS slope is disproportionately influenced by the small-city end of the curve. This makes the OLS slope susceptible to variation in the city size lower threshold and less consistent with our theoretical equations. In SI 2.1, we compare spline-based and OLS-based parameter estimates in greater detail.  

\subsection{Projections}
\label{methods:projections}
The core idea of our projection method is to take the closed-form equations from Methods \ref{methods:theory} and populate them with our spline-based measurements from Methods \ref{methods:measurement}. The logic is that the equations capture the form of the relationship between parameters, while our spline-based measurements provide the most accurate values for these parameters (see SI 2.2).

\begin{table}[t]
    \centering
\begin{tabular*}{0.8\linewidth}{@{\extracolsep{\fill}}lcc}
\\[-1.8ex]\hline\hline \\[-1.8ex]Independent \textbackslash{} Dependent& \multicolumn{2}{c}{Size-growth slope $\beta$} \\
\midrule
Urban population share & -0.049*** & -0.049*** \\
 & (0.004) & (0.012) \\
Country fixed effect & No & Yes \\
Observations & 891 & 891 \\
$R^2$ & 0.168 & 0.545 \\
\hline\hline \\[-1.8ex]& \multicolumn{2}{r}{$^{*}$p$<$0.1; $^{**}$p$<$0.05; $^{***}$p$<$0.01} \\
\end{tabular*}
    \caption{Association between country urbanization and $\beta$. Column (1): pooled specification. Column (2): country fixed effects. The sample is the global cities dataset (Table \ref{tab:1}). }
    \label{tab:2}
\end{table}

We start by projecting $\beta$. We consider four scenarios:
\begin{itemize}
    \item Proportional growth: We set $\beta_{tc} = 0$ for all $t \geq 2020$ and all countries $c$. 
    \item Increasing returns: We set $\beta_{tc} = 0.03$ for all $t \geq 2020$ and all countries $c$. 
    \item Current trend extrapolation: We set $\beta_{tc} = \bar{\beta}_c$ for all $t \geq 2020$ and all countries $c$. Here,  $\bar{\beta}_c$ is the average $\beta$ for country $c$ over 1975-2025 (as mapped in Figure \ref{fig:1}C). 
    \item  Our model: We regress $\beta$ on the urban population share $u$ to obtain $\rho = -0.049$ (Table \ref{tab:2}). Then, for $t \geq 2020$, we set:
    \begin{equation}
        \beta_{tc} = \bar{\beta}_c + \rho \cdot (u_{ct} - \bar{u}_c) \ .
    \end{equation}
    Here, $u_{ct}$ is the projected urban population share for country $c$ from \cite{chen2022updating} (Methods \ref{methods:data}) and $\bar{u}_c$ is the average urban population share for country $c$ over 1975-2025. 
\end{itemize}

We project $\alpha$ for $t \geq 2030$ by plugging the projected $\beta$s into an empirical version of equation \eqref{eq:power_law_change_iterated}:
\begin{equation}
    \alpha_t = \exp\Big(\log(\alpha_{2020}) + \sum_{s \in 2020, \cdots, t - 10} \log(1 + \beta_{s})\Big) \ .
\end{equation}

We project the share of \emph{total} population living in 1M$+$ cities using a two-step approach. 
\begin{enumerate}
    \item First, we project the share of \emph{urban} population in 1M$+$ cities using an empirical version of equation \eqref{eq:shares_vs_rank_size_slope}:
    \begin{equation}
    \label{eq:shares_vs_rank_size_slope_empirical}
        m_t = \frac{x_{t, \max}^{1 - 1/\alpha_t} - z^{1- 1/\alpha_t}}{x_{t, \max}^{1- 1/\alpha_t} - x_{t, \min}^{1- 1/\alpha_t}} \ .
    \end{equation}
    The key challenge in evaluating the right-hand side of this equation is to set plausible values for $x_{t,\min}$ and $x_{t,\max}$, the lower and upper bounds on the sizes of a country's cities. For $x_{t, \min}$, we use the city size lower bound used throughout the paper, $x_{t, \min} = 5000$. For $x_{t, \max}$, the choice is less straightforward because there is no clear upper bound for the size of a country's cities. We assume that $x_{t, \max}$ is proportional to the total urban population of a country $U_{t}$, i.e., $x_{t, \max} = \omega \cdot U_{t}$ where $\omega$ is a tunable parameter. We then calibrate $\omega$ using historical data. For each country, we select $\omega \in (0.1, 2)$ to minimize the average deviation between the shares $m_t$ calculated using equation \eqref{eq:shares_vs_rank_size_slope_empirical} and the shares $\hat{m}_t$ observed directly in the data. Aggregated at the regional level, the calibrated estimates $m_t$ closely match the observed data $\hat{m}_t$ (mean absolute error $|m_t - \hat{m}_t|$ below $0.01$; see SI Fig. S6).
    \item Second, we multiply the projected shares $m_t$ by the projected urban population share $u_t$ to obtain the share of \emph{total} population living in 1M$+$ cities.
\end{enumerate}

\subsection{Code and data availability}
\label{methods:code_and_data_availability}
All data and source code underlying this paper are publicly available. \\

\noindent \textbf{Code}: To enable exact re-execution, we provide a fully automated, end-to-end pipeline that reproduces the entire analysis with a single command. The pipeline is available on GitHub \url{https://github.com/ethz-coss/global-city-growth} and archived on Zenodo at the time of publication \url{https://doi.org/10.5281/zenodo.17339402}.. It handles $\sim 300$ GB of data spanning heterogeneous formats like raster, vector/shapefile, CSV, and Parquet. It executes $100+$ tasks in a directed acyclic graph and produces $200+$ output tables persisted in PostgreSQL (primary store) and processed with DuckDB (for fast processing of large tables). It is containerized with Docker. On a MacBook Pro (Apple M2, 32 GB RAM), a full run completes within $\sim 5$ hours. \\

\noindent \textbf{Source data}: We deposited all redistributable source data at \url{https://doi.org/10.5281/zenodo.17240843}. Some datasets (IPUMS and NHGIS) are public but not redistributable. The pipeline retrieves them via the IPUMS API. Running the pipeline requires a personal IPUMS API key (see \url{https://developer.ipums.org/docs/v2/apiprogram/}). \\

\noindent \textbf{Output data}: A full database dump of all aggregate tables resulting from a pipeline run is available at \url{https://doi.org/10.5281/zenodo.17338967}. The key output datasets --- city boundaries and populations for the USA (1850-2020) and the world (1975-2025) and projections of population shares in 1M$+$ cities by country --- are available at \url{10.5281/zenodo.17315337}. 

\subsection{Acknowledgments}
Andrea Musso and Dirk Helbing acknowledge support from the European Research Council (ERC) under the European Union’s Horizon 2020 research and innovation program (833168). Frank Neffke acknowledges financial support from the Austrian Research Promotion Agency (FFG) in the framework of the project ESSENCSE (873927), within the funding program Complexity Science. Diego Rybski acknowledges support from the German Research Foundation (DFG) project UPon (451083179) and project Gropius (511568027). Andrea Musso is grateful to Simone Daniotti, Cesare Carissimo, and Ricardo Hausmann for helpful conversations. 

\newpage 

\begin{center}
    \LARGE Supplementary Information: Large cities lose their growth advantage as countries urbanize
\end{center}
\vspace{2em}

\setcounter{section}{0}
\setcounter{figure}{0}
\setcounter{table}{0}
\setcounter{equation}{0}

\renewcommand{\thefigure}{S\arabic{figure}}
\renewcommand{\thetable}{S\arabic{table}}

\addtocontents{toc}{\protect\setcounter{tocdepth}{2}}
\tableofcontents
\newpage

\section{Data and definitions}
\subsection{Harmonization of historical US census data}
\begin{figure}[h]
    \centering
    \includegraphics[width=\textwidth]{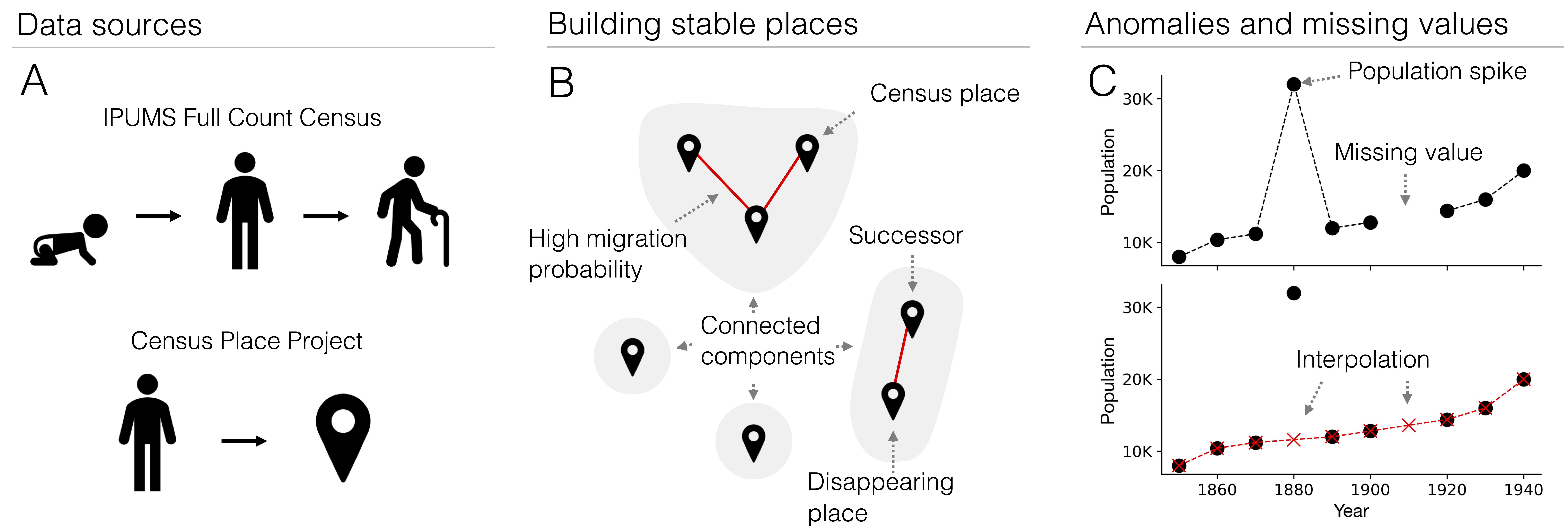}
    \caption{Harmonization of historical US census data. \textbf{(A)} Our census place population estimates build on two data sources: the Census Place Project and the IPUMS full count census. The Census Place Project maps individuals to historical census places, allowing us to infer their population. The IPUMS full count census links individuals across census years, allowing us to compute migration flows between census places. \textbf{(B)} The historical population data contains significant noise, such as census places that disappear or exhibit implausible population spikes. We resolve these issues by constructing a network based on migration flows, where the connected components of the network define geographically and temporally stable places. \textbf{(C)} We use linear interpolation to fix anomalies and impute missing values in the population time series of our stable census places.}
    \label{fig:si1}
\end{figure}

\noindent Our US cities data between 1850 and 1940 comes from two primary data sources: the IPUMS full count census and the Census Place Project (Supp. Fig. \ref{fig:si1}A). 

The IPUMS full count census is the result of a massive digitization project that transformed all US censuses into machine-readable format. These machine-readable censuses contain a wealth of socio-demographic information about every single person who lived on US soil. The key information used here is the census linking key: a unique identifier that tracks individuals across different censuses, making it possible to trace their movements over time.The Census Place Project complements the IPUMS full count census by providing more granular geographic information about the location of individuals. Using a complex matching procedure on the strings of a digitized version of the original handwritten censuses, the authors map each individual in the census to one of almost 70,000 \emph{historical census places}.

Both datasets draw on historical USA decennial census records,
which have been lawfully released to the public. 
The version used in this paper is a public-use, coded dataset that omits direct identifiers such as names and street addresses. Our analysis relies on these lawfully released, de-identified records and, since we use them only to construct aggregate place-level measures, complies with applicable data protection and privacy laws.

Our pipeline uses the Census Place Project to estimate the population of historical census places, and the linking from the IPUMS full count census to denoise these estimates. It features four steps. 

In the first step, we standardize the definition of census place across our data. The historical data (1850-1940) uses ``historical census place'' as its base unit, whereas the modern data (1990-2020) uses ``NHGIS census place''. We match the 70,000 historical census places to 34,000 NHGIS census places by assigning each historical place to its nearest NHGIS neighbor and adopt the NHGIS census place as our base unit. 

In the second step, we use the above matching to estimate the population of NHGIS census places during the historical period (1850-1940). These estimates have two key issues: (i) there are a non-negligible number of census places (around 17\% of the total) that disappear: they have a non-zero population in some year $t_e$ and zero population in years $t > t_e$;
(ii) there are population spikes: given three consecutive observation years $t_1, t_2$ and $t_3$, the population of the census place doubles between years $t_1$ and $t_2$ and then halves between $t_2$ and $t_3$. This noise is not surprising. The data come from handwritten documents that are over one hundred years old. 

In the third step, we address this noise by re-allocating population across census places. To support our re-allocation decisions, we estimate migration flows between places using the census linking. This unique identifier allows us to follow the movement of individuals across different censuses and hence estimate the probability of migration between places. We use these migration probabilities to match disappearing places with successors. For each disappearing place, we identify its successor as the location within a 50 km radius that has the highest probability of receiving its migrants. If migration data is unavailable, we simply choose the nearest place. 

This process results in a network where nodes are census places and edges link each disappearing place to its assigned successor (Fig. \ref{fig:si1}B). We enrich this network by adding an edge between two places if the migration probability between them is ``high'', where high means that two places within 50 km distance have a migration probability exceeding 0.5 (50\% of the people living in one place moved to the other between two observations).

If two places in the network are linked by an edge, we consider them the same place. So, the connected components of the network identify unique places. We assign each connected component a unique identifier using the NHGIS ID of the member of the component with the highest total population. We then estimate the population of each connected component by summing the populations of its members. This results in a dataset of 30,000 census places that are stable throughout our historical period. 

In the fourth and final step, we clean the population time series of our stable census places. First, we apply heuristic rules to flag implausible data points, such as extreme population spikes (Fig. \ref{fig:si1}C). Then we use linear interpolation to impute these flagged data points and existing missing values. The result is a harmonized panel dataset covering the populations of 30,000 census places from 1850 to 1940. The combination of this dataset and the much cleaner modern version (1990-2020) powers our longitudinal analysis of the US.  

\subsection{A visual comparison of administrative, functional, and geographic city definitions}
\begin{figure}[h!]
    \centering
    \includegraphics[width=0.9\textwidth]{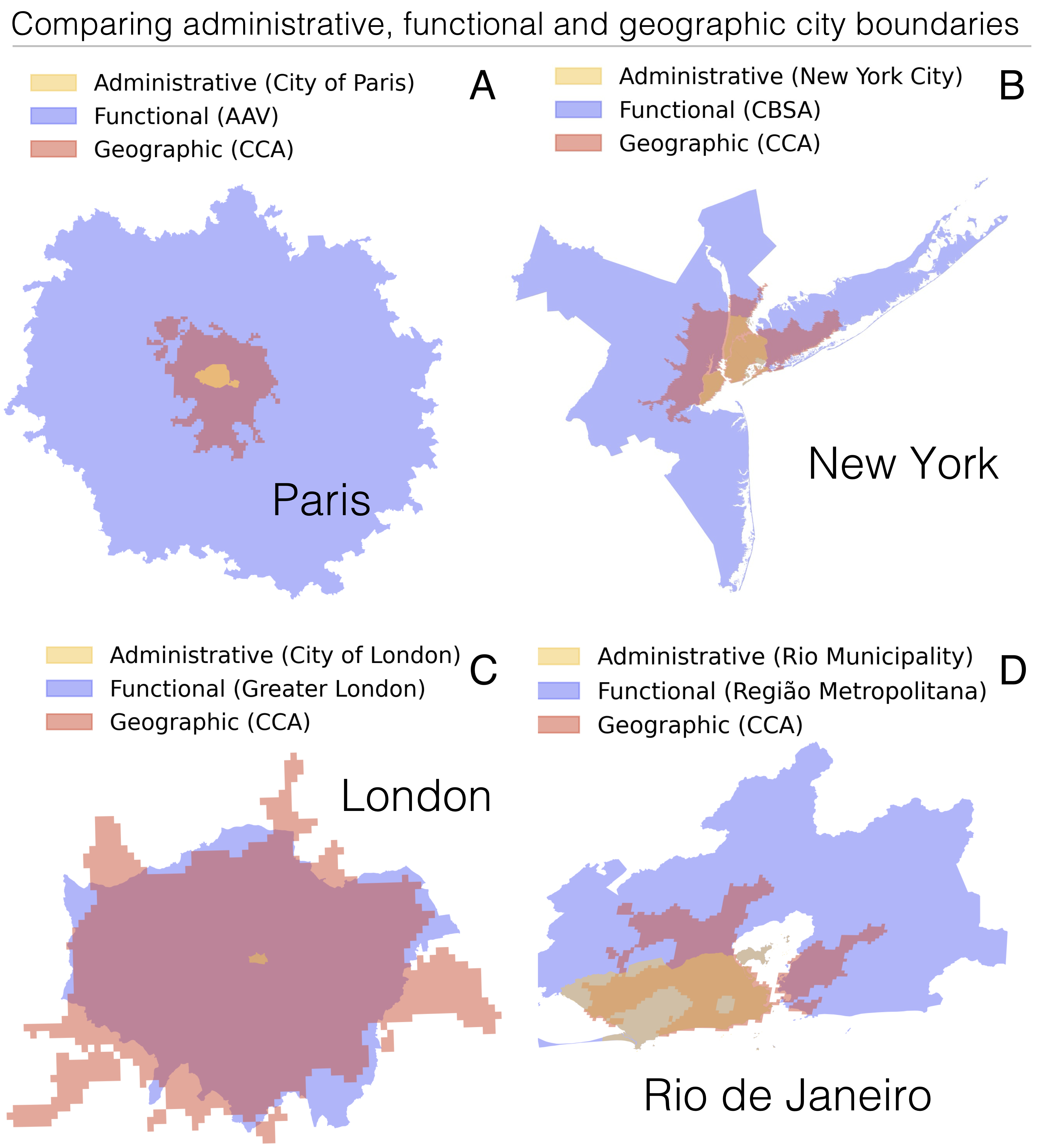}
    \caption{A comparison of administrative, functional, and geographic city definitions for (A) Paris, (B) New York, (C) London, and (D) Rio de Janeiro. The geographic definition (red) is the one used in this study, derived by applying the City Clustering Algorithm (CCA) to the grids from the Global Human Settlement Layer. The administrative (orange) and functional (blue) boundaries are based on the following official definitions: \textbf{(A)} Paris: The administrative area is the City of Paris (the 20 arrondissements); the functional area is the Aire d'attraction d'une ville (AAV), which is based on commuting patterns. \textbf{(B)} New York: The administrative area is New York City (the five boroughs); the functional area is the Core-Based Statistical Area (CBSA), which is based on economic and commuting ties. \textbf{(C)} London: The administrative boundary is the historic City of London; the functional boundary is the Greater London area, an administrative region composed of the City of London and 32 surrounding boroughs, which largely coincides with the area defined by commuting patterns. \textbf{(D)} Rio de Janeiro: The administrative boundary is the Rio Municipality; the functional boundary is the Região Metropolitana, an area of socio-economically integrated municipalities defined by state law for the joint planning of public services.}
    \label{fig:si2}
\end{figure}
Broadly speaking, city definitions fall into three categories: administrative, functional, and geographic. In Figure \ref{fig:si2}, we illustrate these different definitions for four large cities: Paris, New York, London, and Rio de Janeiro. 

Administrative definitions are based on political boundaries. Their main strength is that they are widely available. From small towns to sprawling megacities, most settlements have well-defined political boundaries. However, these boundaries are rigid or slow to change, leading to a distorted picture of a city’s actual growth. For example, all four political cities (orange) in Figure \ref{fig:si2} are far smaller than their urban areas (red), indicating that these cities have spilled over their political boundaries. 

Functional definitions (like the widely used Metropolitan Statistical Area (MSA)) are based on commuting patterns. These definitions often encompass the urban area itself and a wider region surrounding it from where people commute for work (see blue areas in Figure \ref{fig:si2}).  These areas provide the most accurate city definition from an economic perspective because they capture the gravitational pull of the city beyond its geographical boundaries. The core weakness of functional definitions lies in their coverage and comparability. Functional urban areas are typically constructed by hand on a case-by-case basis, limiting their availability to large-ish cities. Additionally, criteria for defining these areas change across countries and over time, limiting comparability.

\begin{figure}[h!]
    \centering
    \includegraphics[width=0.9\textwidth]{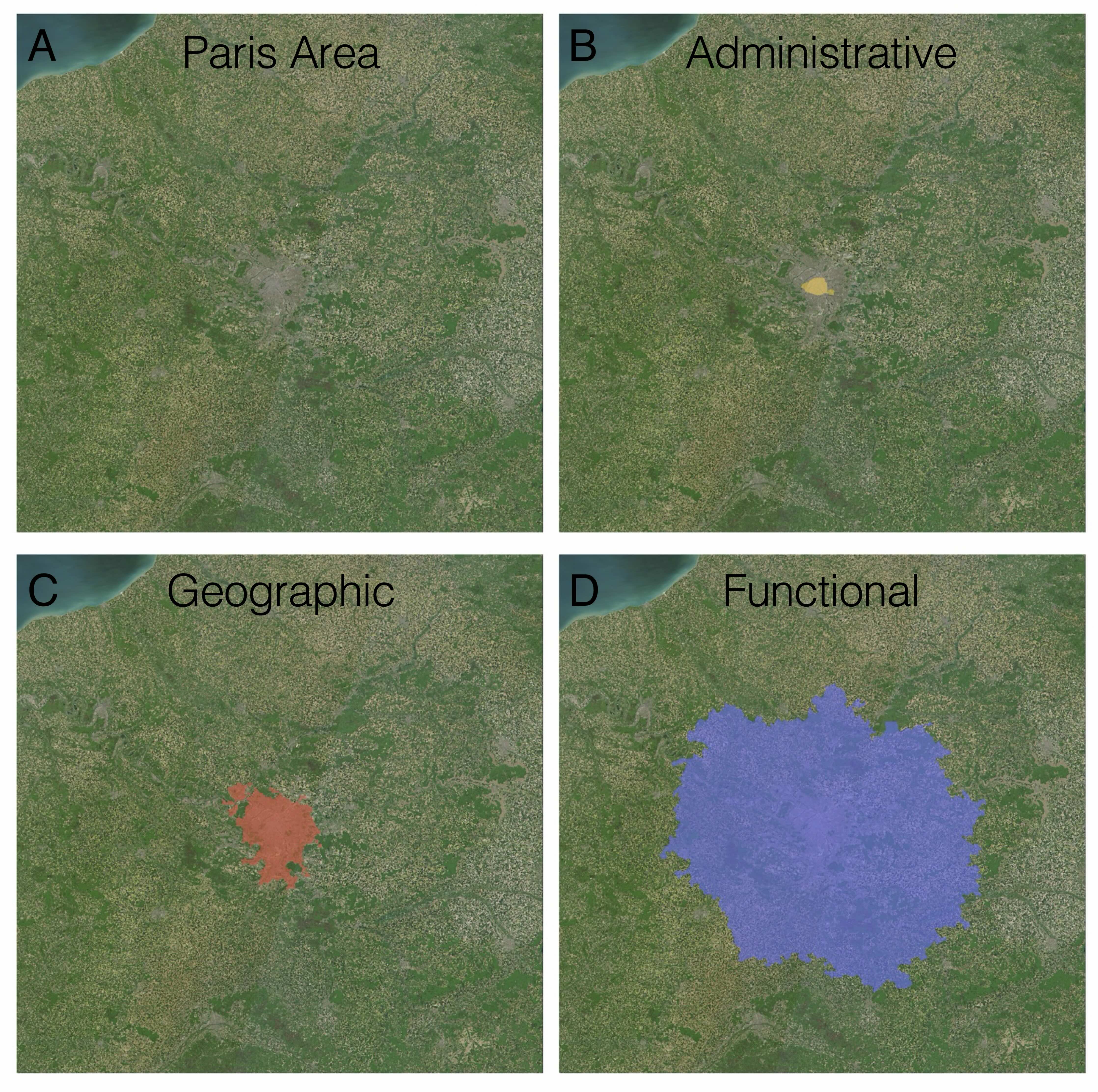}
    \caption{Three distinct definitions of ``Paris'' against a satellite image of northern France. The administrative definition (orange) covers just the core of the continuous built-up area. The geographic definition (red) roughly coincides with this built-up area. The functional definition (blue) extends significantly further, encompassing a broad region of northern France that includes numerous surrounding rural areas with close economic or social connections to the urban center.}
    \label{fig:si3}
\end{figure}

Geographic definitions are based on built-up area or population density thresholds. The city boundaries issued from these definitions tend to coincide with those of the ``urban area'', as defined by different statistical agencies. Intuitively, they coincide with the agglomeration of built-up area (Fig. \ref{fig:si3}), and so their boundaries lie somewhere in the middle between the broader functional urban area and the smaller administrative area (Fig. \ref{fig:si2}). Their disadvantage is that they fail to capture the gravitational pull of the city beyond its built-up area. Their advantages are that they adapt naturally as cities expand, and they are both widely available and highly comparable. In fact, advances in satellite imagery and algorithmic processing make it possible to track an unprecedented range of settlements over long time windows, all while keeping similar city definitions. While no definition is perfect, our view is that geographic definitions offer the best balance between coverage, comparability, and adaptation for studying city growth on a large spatial and temporal scale.

\newpage
\section{Methodology}
\subsection{Spline-based vs OLS-based parameter estimates}
In this section, we explain why we prefer using spline-based methods over traditional linear models for measuring the parameters $a$ and $b$ of our theoretical model. Generally, we want our empirical estimates $\alpha$, $\beta$ for $a$, $b$ to satisfy three desirable properties:
\begin{enumerate}
    \item[(i)] $\alpha$ and $\beta$ coincide with $a$ and $b$ when the model assumptions hold exactly, i.e., when the rank-size and size-growth curves are straight lines in log-log space. 
    \item[(ii)] $\alpha$ and $\beta$ are robust to arbitrary empirical design choices, such as the city size lower threshold.
    \item[(iii)] $\alpha$ and $\beta$ are consistent with the equations derived from the theoretical model (e.g., $\alpha_{t+10} = \alpha_t \cdot ( 1 + \beta_t)$) even when the empirical data deviate from the perfect linearity assumed by the model. 
\end{enumerate}

\begin{figure}[h]
    \centering
    \includegraphics[width=0.95\textwidth]{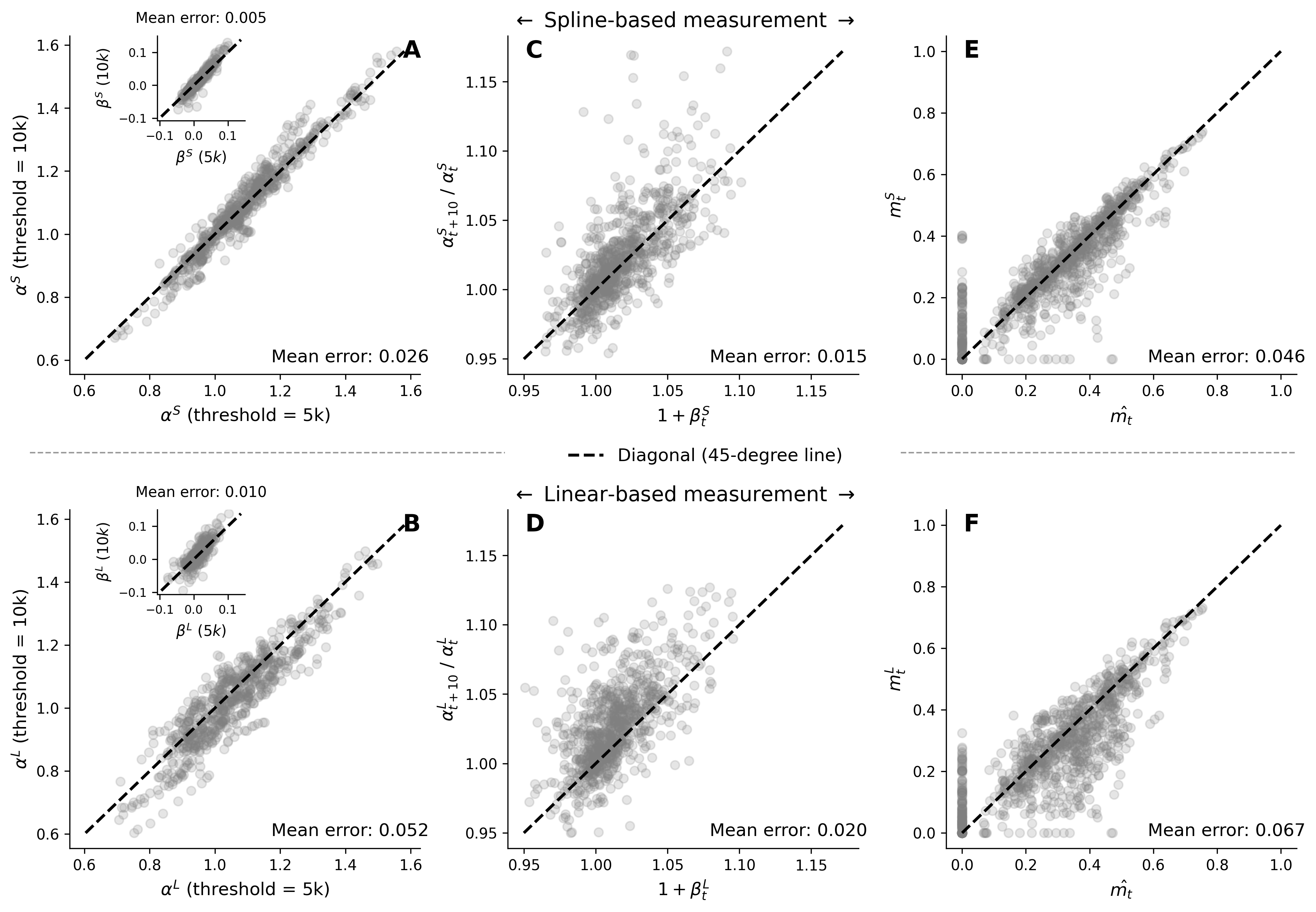}
    \caption{Comparison of spline-based and OLS-based parameter estimates. \textbf{(A-B)} Scatterplots showing $\alpha$ and $\beta$ estimated with a city size threshold at 5,000 vs. 10,000 people for (A) spline-based and (B) OLS-based estimates. \textbf{(C-D)} Scatterplots of $\alpha_{t+10} / \alpha_t$ against $1 + \beta_t$ for (C) spline-based and (D) OLS-based estimates. For visual clarity, we omit the top and bottom $1\%$ of data points so that extreme values do not compress the axes and obscure the central pattern. \textbf{(E-F)} Scatterplots of model shares $m_t$ vs. observed shares $\hat{m}_t$ for (E) spline-based and (F) OLS-based estimates. }
    \label{fig:si4}
\end{figure}

Our spline-based $\alpha^S, \beta^S$ satisfy these three properties more closely than OLS-based $\alpha^L, \beta^L$. Property (i) is clearly satisfied by both. For property (ii), Figure \ref{fig:si4}A-B shows that $\alpha^S$, $\beta^S$ are more robust to a change in city size threshold from 5,000 to 10,000 people. For property (iii), Figure \ref{fig:si4}C-F shows that our two key equations hold more closely with spline-based estimates. For the equation relating $a$ and $b$, points align more clearly on the 45-degree line and have a 25\% lower mean absolute error $|\alpha_{t+10} / \alpha_t - (1 + \beta_t)|$ (Fig. \ref{fig:si4}C-D). For the equation relating $a$ and $m$ (the share of urban population living in 1M$+$), the spline-based estimate $m^S_t$ fits the empirical data $\hat{m}_t$ better, reducing the mean absolute error by approximately 30\% compared to the OLS-based estimate $m^{L}_t$ (Fig. \ref{fig:si4}E-F). 

We speculate that the greater robustness of our spline-based estimates has to do with weighting. Our spline-based slopes summarize a curve by giving \emph{equal weight} to each part of the x-range. In contrast, the OLS slope summarizes a curve by giving different weights to different parts of the range, based on the density of data points within them. Put differently, the OLS slope is a \emph{data-weighted} average of many local slopes that places more weight where observations are more dense. In city population data, small cities are far more numerous than large ones, so the OLS slope is dominated by outcomes in the small-city end of the curve. As a result, the OLS slope summarizes the sample composition more than the shape of the relationship. This makes the OLS slope less robust to changes in the city size lower threshold, because small cities are weighted so heavily. It also makes OLS slopes less consistent with our theoretical equations, because these describe a curve’s geometry independent of how densely observations are distributed along it.

\begin{figure}[h]
    \centering
    \includegraphics[width=0.95\textwidth]{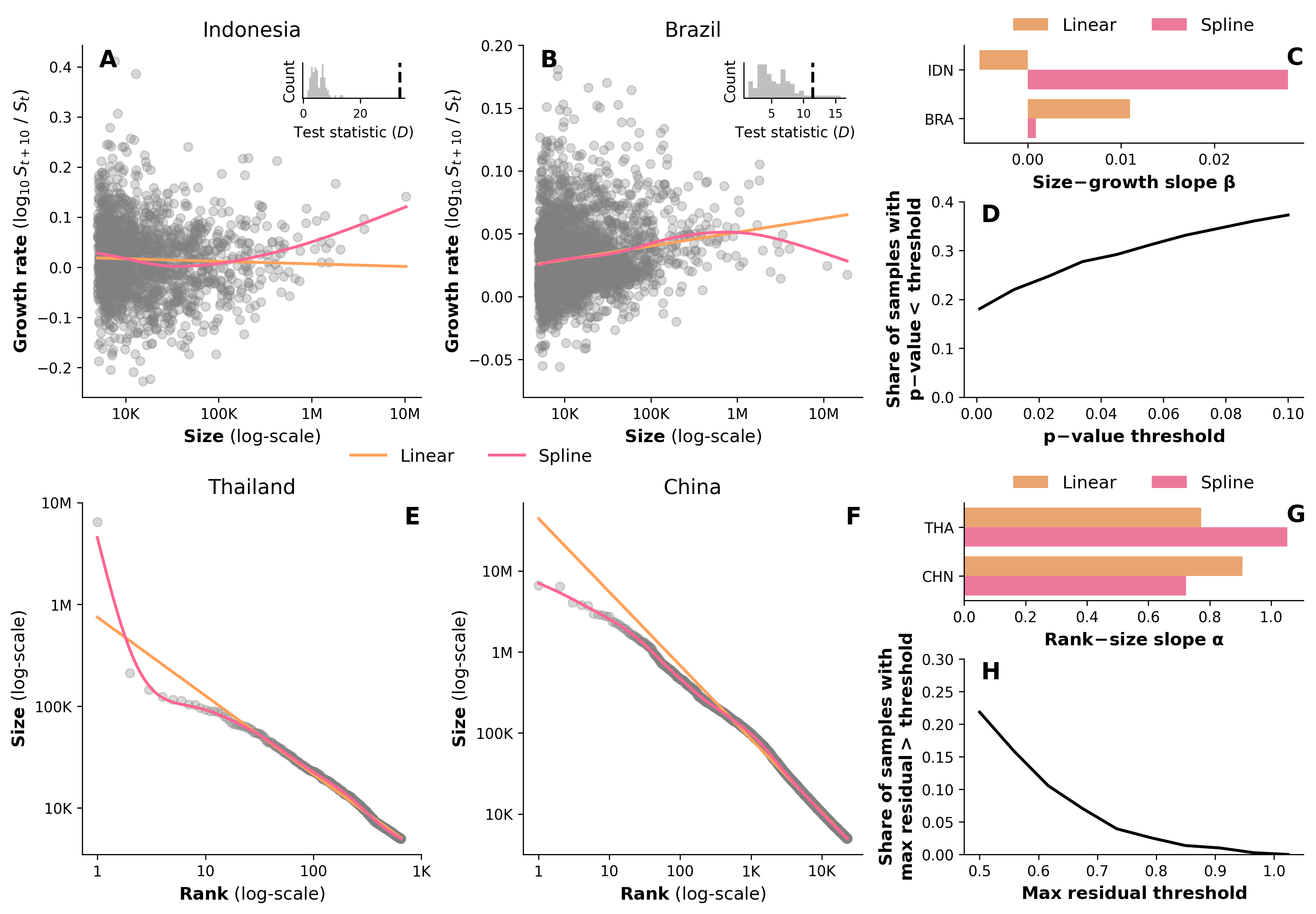}
    \caption{Examples of OLS slopes poorly fitting size-growth and rank-size curves. \textbf{(A-B)} Size-growth curves for Indonesia and Brazil (pink = penalized B-spline fit with $\lambda = 100$, orange = OLS fit, grey = data points). Insets: the distribution (grey) and test statistic (dashed black line) of the log-likelihood ratio test described in the text. \textbf{(C)} Slope of the size-growth relationship ($\beta$) from an OLS-based/spline-based estimate. \textbf{(D)} Results of the log-likelihood ratio test by p-value threshold: approximately 30\% of country-years reject linearity at $p < 0.05$. \textbf{(E-F)} Rank-size curves for Thailand and China  (pink = penalized B-spline fit with $\lambda = 1$, orange = OLS fit, grey = data points). \textbf{(G)} Slope of the rank-size relationship ($\alpha$) from an OLS-based/spline-based estimate. \textbf{(H)} Share of country-years with maximum-residual above threshold: more than 20\% of samples have a maximum residual greater than 0.5.}
    \label{fig:si5}
\end{figure}

Figure \ref{fig:si5}A-B shows examples in which OLS slopes provide poor estimates for the shape of the size-growth curve. The figure highlights a U-shaped curve in Indonesia, meaning that small and large cities grow faster than middle-sized ones. It also highlights an inverse-U shaped curve in Brazil, meaning that small and large cities grow slower than middle-sized ones. When we fit these patterns with an OLS, the slope only reflects what is happening at the left end of the data (small to medium size). As a result, the OLS slope provides a poor summary of the full curve. In Indonesia, large cities are growing much faster than the rest, but the OLS slope is negative. Conversely, in Brazil, large cities are growing slower, but the OLS slope is positive (Fig. \ref{fig:si5}C). 

Bent curves are quite common, so this problem shows up in a non-negligible number of cases. We can see this through a simple log-likelihood ratio test. Our null hypothesis $H_0$ is that the relationship between log-size and log-growth in a given country and year is linear. Our alternative hypothesis $H_1$ is that it is not linear. Our test statistic is the log-likelihood ratio, $D = 2 \cdot (l_{\text{spline}} - l_{\text{linear}})$, which measures the improvement in fit offered by the spline model with respect to a linear one. If $D$ is large, we know that the spline model greatly improves the fit, suggesting that the data present some important non-linearities, and hence that the linear model is likely misspecified. 

To determine what ``large'' means for our test statistic $D$, we estimate it under the null hypothesis (linearity) using a wild bootstrap procedure. This procedure involves the following steps:
\begin{enumerate}
    \item We fit a linear model to the original data using OLS regression.
    \item We generate a synthetic dependent variable by adding the linear model's predicted values $\hat{y}$ to its residuals $y - \hat{y}$ multiplied by a random sign $w$:
    \begin{equation}
        y_{\text{boot}} = \hat{y} + w \cdot (y - \hat{y}) \, \quad \ w = \text{random weights, -1 or +1 with probability 0.5} \ .
    \end{equation}
    \item We estimate the linear and spline model with this synthetic dependent variable to obtain the log-likelihoods $l^{\text{boot}}_{\text{linear}}$ and $l^{\text{boot}}_{\text{spline}}$ and the bootstrapped test statistic $D_{\text{boot}} = 2 \cdot (l^{\text{boot}}_{\text{spline}} - l^{\text{boot}}_{\text{linear}})$.
    \item We repeat steps 2 and 3 for $N = 1000$ iterations to obtain a distribution of plausible values of our test statistic under the null-hypothesis.
\end{enumerate}
We define our test p-value by $p = P(D \leq D_{\text{boot}})$, i.e., the probability that the true test statistic $D$ is smaller than the test statistics generated under the null hypothesis. If $p < 0.05$ we know that our true statistic is larger than $95\%$ of the values generated under the null hypothesis --- $D$ is ``large'' --- and hence that the data presents non-linearities (Fig. \ref{fig:si5}D). The insets of Figure \ref{fig:si5}A-B show the observed test statistic and the bootstrapped distribution for Indonesia and Brazil.

Figure \ref{fig:si5}E-F shows examples in which OLS slopes provide poor estimates for the shape of the rank-size curves. We observe a common example of non-linearity: urban primacy. Bangkok --- Thailand's largest city --- is over 30 times larger than the next city and home to more than 50\% of Thailand's urban population. So, as we move from rank 1 to rank 2 in the Thai urban system, the slope of the rank-size curve declines far faster than when moving between higher ranks. The OLS slope fails to capture this pattern. Skewed by the many small cities, the slope largely ignores the outsized Bangkok. The result is a poor summary of Thailand's rank-size curve. Intuitively, Thailand has a very steep rank-size curve --- the majority of people live in the biggest city --- but the OLS slope suggests the opposite (Figure \ref{fig:si5}G).

Thailand is one of the most extreme cases, but non-linear rank-size curves are fairly common (e.g., China in Fig. \ref{fig:si5}F). To see this, we fit an OLS regression to log-rank vs. log-size data points, and then consider the maximum residual of this linear fit. The larger this maximum residual, the worse the curve fits the data. We observe that maximum residuals vary between 0.05 (very small) and 1.1 (large), with about 20\% of the sample having a maximum residual above 0.5 (Fig. \ref{fig:si5}H). 

\subsection{Projections}
\begin{figure}[h]
    \centering
    \includegraphics[width=0.95\textwidth]{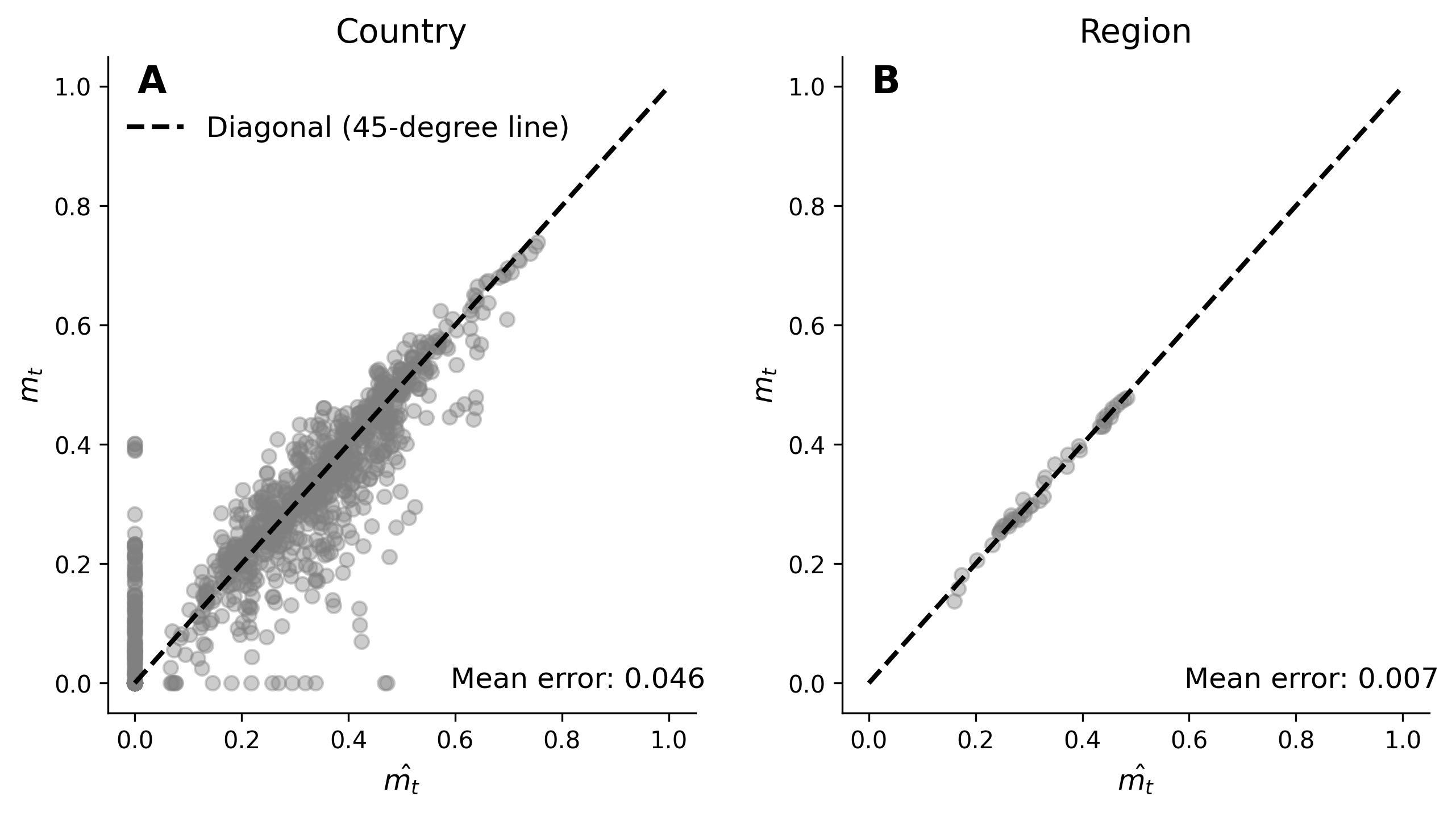}
    \caption{Relationship between the modeled share of the urban population living in cities with at least one million people, $m_t$, and the observed share $\hat{m_t}$. The scatterplots show $m_t$ vs. $\hat{m_t}$ at the \textbf{(A)} country level and \textbf{(B)} regional level. In both panels, points cluster along the diagonal (45-degree) line, indicating a good fit.}
    \label{fig:si6}
\end{figure}

In this section, we explain how we calibrate the model to project the share of the urban population living in cities above one million, $m_t$. We start from the equation:
\begin{equation}
\label{eq:si_shares_vs_rank_size_slope}
    m_t = \frac{x_{t, \max}^{1 -1/\alpha_t} - z^{1 -1/\alpha_t}}{x_{t, \max}^{1 - 1/\alpha_t} - x_{t, \min}^{1 - 1/\alpha_t}} \ .
\end{equation}
Herein, $\alpha_t$ is the empirical rank-size slope, $x_{t, \min}$ and $x_{t, \max}$ are the lower and upper bounds on the sizes of a country's cities, and $z = \text{1 million}$. As noted in the main text, the key challenge in evaluating equation \eqref{eq:si_shares_vs_rank_size_slope} is to set the upper bound $x_{t, \max}$. 

Our approach is to assume that $x_{t, \max}$ is proportional to the total urban population of a country $U_t$, i.e., $x_{t, \max} = \omega \cdot U_t$ where $\omega$ is calibrated using historical data. For each country, we grid-search $\omega \in (0.1, 2)$ and choose the value that minimizes the maximum absolute deviation between our model's estimates $m_t$ and the observed share $\hat{m}_t$ during the period 1975-2025, i.e., $\omega = \arg \min_{\omega \in (0.1,2)} ( \max_{t \in [1975, 2025]}|\hat{m}_t - m_t(\omega)| )$. Figure \ref{fig:si6}A shows that this calibration yields a strong fit between the estimated and observed share, with a mean absolute error $|m_t - \hat{m}_t|$ of 0.046. 

For regional analysis, we calculate the aggregate share $m_{rt}$ for a region $r$ by taking the weighted average of country-level shares, where weights are based on each country's total urban population $U_{ct}$:
\begin{equation}
    m_{rt} = \sum_{c \in r}\frac{U_{ct}}{\sum_{c \in r} U_{ct}} m_{ct}  \ .
\end{equation}
At the regional level, the correlation between our estimated shares and the observed data is even stronger, with a mean absolute error below 0.01 (Fig. \ref{fig:si6}B).

\subsection{Deriving interaction between rank-size slope and scaling exponent}
\label{methods:derivations:scaling}
Here, we derive an equation that describes how city-level scaling laws interact in the aggregate with the distribution of population (see also \cite{gomez-lievano2012statistics}). We begin with two assumptions:
\begin{enumerate}
    \item A country has $N$ cities, and their sizes (or populations) $S_i$ follow a power-law distribution with exponent $\alpha$, meaning that the share of cities larger than size $x$ is proportional to $x^{-1/\alpha}$
    \item An outcome $Y$ scales super-linearly with city population: $Y_i \propto S_i^{\theta}$, where $\theta > 1$ is the urban scaling exponent. 
\end{enumerate}
From these assumptions, our goal is to relate the total urban population $U$ to the aggregate value of the outcome $Y_{\text{agg}}$:
\begin{equation}
    U = \sum_{i = 1}^N S_i \ \quad \ Y_{\text{agg}} = \sum_{i= 1}^N Y_i = \sum_{i}^NS_i^{\theta}
\end{equation}
We do so by analyzing the scaling behavior of these two sums with the number of cities $N$, as described by two important statistical theorems:
\begin{enumerate}
    \item Law of Large Numbers (LLN): A sum of power-law distributed random variables with exponent $\alpha$ \emph{smaller} than one scales linearly with the number of terms.
    \item Generalized Central Limit Theorem (GCLT): A sum of power-law distributed random variables with exponent $\alpha$ \emph{greater} than one is dominated by the largest term and scales as $N^{\alpha}$.
\end{enumerate}
We start by noting that if the city size distribution follows a power-law with exponent $\alpha$, then the outcome $Y$ follows a power-law distribution with exponent $\alpha \theta$. From this observation, and the two theorems above, we deduce three distinct regimes are possible depending on whether these exponents are greater or less than $1$.
\begin{enumerate}
    \item When $\alpha < 1$ and $\alpha \theta < 1$, the LLN applies to both sums:
\begin{equation}
     Y_{\text{agg}} \propto N \ , \quad \ U \propto N \ \quad \ \Rightarrow \ \quad \ Y_{\text{agg}} \propto U \ .
 \end{equation}
 \item When $\alpha < 1$ and $\alpha \theta > 1$, the LLN applies to $U$, while the GCLT applies to $Y_{\text{agg}}$:
 \begin{equation}
    Y_{\text{agg}} \propto N^{\alpha \theta} \ , \quad \ U \propto N \ \quad \ \Rightarrow \ \quad \ Y_{\text{agg}} \propto U^{\alpha \theta} \ .
 \end{equation}
 \item When $\alpha > 1$ and $\alpha \theta > 1$, the GCLT applies to both sums:
 \begin{equation}
    Y_{\text{agg}} \propto N^{\alpha \theta} \ , \quad \ U \propto N^{\alpha} \ \quad \ \Rightarrow \ \quad \ Y_{\text{agg}} \propto U^{\theta} \ .
 \end{equation}
\end{enumerate}
These asymptotic results show that more top-heaviness (a larger $\alpha$) amplifies aggregate outcomes, with the effect being particularly strong in the intermediate regime $\alpha < 1$ and $\alpha \theta > 1$. The apparent vanishing of this amplification at extreme values of $\alpha$ is an artifact of this asymptotic approach. In any finite system, concentrating population always amplifies outcomes due to the convexity of the scaling function. Thus, for a constant $\theta$ (but see \cite{ribeiro2021association}), the amplifying effect of concentration is better understood as following an inverse-U shape: it is strongest in the intermediate regime and diminishes, but remains positive, at the extremes.

\newpage
\section{Hyperparameter robustness}
In this section, we discuss the robustness of our results to changes in the hyperparameters used to conduct our analysis. We discuss robustness separately for the global and US data. 

\subsection{Global data}
In the global data, our key hyperparameters are the degree-of-urbanization threshold, the city population threshold, and the country filtering conditions. To analyze robustness, we replicate Table 2 from the main text, which summarizes succinctly the core results underpinning most of the paper's conclusions: the weakening growth advantage of large cities. 

\subsubsection{Degree-of-urbanization threshold}
The degree-of-urbanization threshold sets the minimum level of urbanity for a grid cell to be considered ``urban''. In the smod dataset of the GHSL, each cell is assigned a degree-of-urbanization imputed from built-up area estimates derived from satellite imagery and population estimates derived from censuses.  The GHSL defines the following degrees-of-urbanization: 
\begin{enumerate}
    \item[30] Urban centre  
    \item[23] Dense urban cluster
    \item[22] Semi-dense urban cluster
    \item[21] Suburban
    \item[13] Rural cluster
    \item[12] Low density rural
    \item[11] Very low density
\end{enumerate}

\begin{table}[h!]
    \centering
    \begin{tabular*}{0.8\linewidth}{@{\extracolsep{\fill}}lcc}
    \\[-1.8ex]\hline\hline \\[-1.8ex]Independent \textbackslash{} Dependent& \multicolumn{2}{c}{Size-growth slope} \\
    \midrule
    Urban population share & -0.055*** & -0.043*** \\
     & (0.004) & (0.012) \\
    Country fixed effect & No & Yes \\
    Observations & 810 & 810 \\
    $R^2$ & 0.199 & 0.572 \\
    \hline\hline \\[-1.8ex] Degree-of-urbanization threshold 22 &  \\
    \end{tabular*}
    \bigskip
    \begin{tabular*}{\linewidth}{@{\extracolsep{\fill}}lcc}
    \\[-1.8ex]\hline\hline \\[-1.8ex]Independent \textbackslash{} Dependent& \multicolumn{2}{c}{Size-growth slope} \\
    \midrule
    Urban population share & -0.084*** & -0.091*** \\
     & (0.009) & (0.033) \\
    Country fixed effect & No & Yes \\
    Observations & 207 & 207 \\
    $R^2$ & 0.281 & 0.441 \\
    \hline\hline \\[-1.8ex] Degree-of-urbanization threshold 23 & \multicolumn{2}{r}{$^{*}$p$<$0.1; $^{**}$p$<$0.05; $^{***}$p$<$0.01} \\
    \end{tabular*}
    \caption{Robustness of results to variations in the degree-of-urbanization hyperparameter.}
    \label{tab:rob1}
\end{table}
In the main paper, we set the degree-of-urbanization threshold to semi-dense urban cluster or more $(\geq 22)$. This choice is dictated by the desire to balance two opposing requirements. On the one hand, we want to include as many cities as possible, and have a size spectrum that is as broad as possible. On the other hand, we want to ensure that distinct cities remain separate entities. A smaller threshold, such as 21 (``Suburban''), causes cities in densely populated regions (e.g., the Ganges and Yangtze river deltas) to merge into a single, massive agglomeration. Stricter thresholds, such as 23 (``Dense urban cluster'') or 30 (``Urban Centre''), lead to the exclusion of small cities. Our exploration of the data suggests a threshold at 22 strikes the best balance.  However, our results are robust to other threshold choices. Table \ref{tab:rob1} shows that the direction and statistical significance of our findings remain unchanged with thresholds of 23 and 30. 

\subsubsection{City population threshold}
The city population threshold dictates the minimum population required for a contiguous cluster of urban cells to be defined as a ``city''. In the main paper, we chose 5,000 as the value of this threshold. This choice is motivated by two factors. First, it aligns with the GHSL methodology, which begins its urban classification at this population level. Second, empirical analysis of the city size distribution shows a structural break around 5,000 inhabitants, suggesting data for smaller settlements may be incomplete.

\begin{table}[h!]
    \centering
    \begin{tabular*}{0.8\linewidth}{@{\extracolsep{\fill}}lcc}
    \\[-1.8ex]\hline\hline \\[-1.8ex]Independent \textbackslash{} Dependent& \multicolumn{2}{c}{Size-growth slope} \\
    \midrule
    Urban population share & -0.040*** & -0.023 \\
     & (0.005) & (0.015) \\
    Country fixed effect & No & Yes \\
    Observations & 576 & 576 \\
    $R^2$ & 0.099 & 0.471 \\
    \hline\hline \\[-1.8ex] City population threshold 10,000 &  \\
    \end{tabular*}
    \bigskip
    \begin{tabular*}{\linewidth}{@{\extracolsep{\fill}}lcc}
    \\[-1.8ex]\hline\hline \\[-1.8ex]Independent \textbackslash{} Dependent& \multicolumn{2}{c}{Size-growth slope} \\
    \midrule
    Urban population share & -0.044*** & -0.061*** \\
     & (0.005) & (0.014) \\
    Country fixed effect & No & Yes \\
    Observations & 567 & 567 \\
    $R^2$ & 0.130 & 0.526 \\
    \hline\hline \\[-1.8ex] City population threshold 10,000 (No Angola)&  \\
    \end{tabular*}
    \bigskip
    \begin{tabular*}{\linewidth}{@{\extracolsep{\fill}}lcc}
    \\[-1.8ex]\hline\hline \\[-1.8ex]Independent \textbackslash{} Dependent& \multicolumn{2}{c}{Size-growth slope} \\
    \midrule
    Urban population share & -0.091*** & -0.074** \\
     & (0.012) & (0.036) \\
    Country fixed effect & No & Yes \\
    Observations & 81 & 81 \\
    $R^2$ & 0.439 & 0.617 \\
    \hline\hline \\[-1.8ex] City population threshold 100,000 & \multicolumn{2}{r}{$^{*}$p$<$0.1; $^{**}$p$<$0.05; $^{***}$p$<$0.01} \\
    \end{tabular*}
    \caption{Robustness of results to variations in the city population threshold.}
    \label{tab:rob2}
\end{table}

The results are robust to changes in this threshold. Table \ref{tab:rob2} shows that using population thresholds at 10,000 and 100,000 yields negatively signed coefficients similar to those in the main text. All coefficients are significant, except for that of the regression with country fixed effects at a threshold of 10,000. This loss of significance is due to a single outlier (Angola). Excluding Angola re-establishes the significance of the coefficient.

\subsubsection{Country filtering conditions}
We apply two filtering rules to determine countries to include in our analysis. The first is a minimum threshold for the number of cities in a country (which we set to 50), and the second is the manual exclusion of countries with data quality issues (Nepal and Myanmar).

The minimum threshold for the number of cities in a country is chosen to ensure that our analysis of national city size distributions is meaningful. The growth advantage of large cities and the top-heaviness of the city size distribution have little relevance in countries with only a handful of cities (e.g., Singapore). The choice of 50 is a reasonable, albeit arbitrary, lower bound. Nepal and Myanmar were excluded due to data quality issues observed in our visual exploration of the data.

\begin{table}[h!]
    \centering
    \begin{tabular*}{0.8\linewidth}{@{\extracolsep{\fill}}lcc}
    \\[-1.8ex]\hline\hline \\[-1.8ex]Independent \textbackslash{} Dependent& \multicolumn{2}{c}{Size-growth slope} \\
    \midrule
    Urban population share & -0.050*** & -0.052*** \\
     & (0.003) & (0.011) \\
    Country fixed effect & No & Yes \\
    Observations & 999 & 999 \\
    $R^2$ & 0.171 & 0.549 \\
    \hline\hline \\[-1.8ex] Countries with at least 30 cities &\\
    \end{tabular*}
    \bigskip
    \begin{tabular*}{\linewidth}{@{\extracolsep{\fill}}lcc}
    \\[-1.8ex]\hline\hline \\[-1.8ex]Independent \textbackslash{} Dependent& \multicolumn{2}{c}{Size-growth slope} \\
    \midrule
    Urban population share & -0.046*** & -0.055*** \\
     & (0.004) & (0.011) \\
    Country fixed effect & No & Yes \\
    Observations & 594 & 594 \\
    $R^2$ & 0.188 & 0.605 \\
    \hline\hline \\[-1.8ex] Countries with at least 100 cities & \\
    \end{tabular*}
    \bigskip 
    \begin{tabular*}{\linewidth}{@{\extracolsep{\fill}}lcc}
    \\[-1.8ex]\hline\hline \\[-1.8ex]Independent \textbackslash{} Dependent& \multicolumn{2}{c}{Size-growth slope} \\
    \midrule
    Urban population share & -0.041*** & -0.037*** \\
     & (0.004) & (0.013) \\
    Country fixed effect & No & Yes \\
    Observations & 909 & 909 \\
    $R^2$ & 0.102 & 0.528 \\
    \hline\hline \\[-1.8ex] Base sample $+$ Myanmar and Nepal & \multicolumn{2}{r}{$^{*}$p$<$0.1; $^{**}$p$<$0.05; $^{***}$p$<$0.01} \\
    \end{tabular*}
    \caption{Robustness of results to variations in the sample of countries.}
    \label{tab:rob3}
\end{table}

Our findings are robust to both filtering criteria. Specifically, Table \ref{tab:rob3} shows that our analysis replicates with a minimum threshold for the number of cities at 30 and 100, as well as including Nepal and Myanmar in the sample.

\subsection{US data}
\begin{figure}[h]
    \centering
    \includegraphics[width=0.95\textwidth]{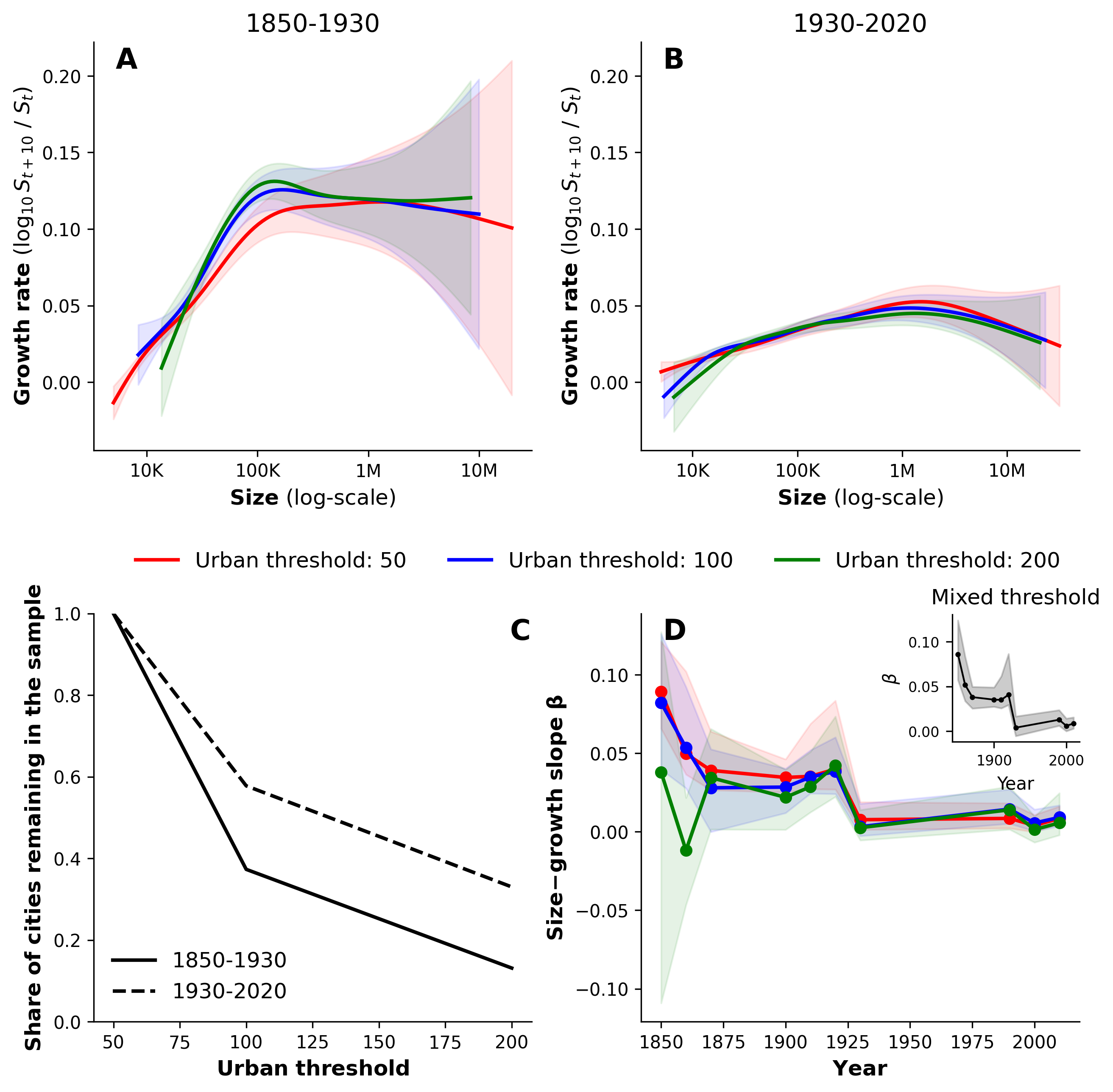}
    \caption{Robustness to variation in the urban threshold for US data. \textbf{(A-B)} The shape of the size-growth curve for different urban threshold choices. \textbf{(C)} Share of cities remaining in the sample by threshold: raising the threshold from 50 to 200 removes over 80\% of cities in the early period (1850-2020). This helps explain the instability of the green curve in (D). \textbf{(D)} Historical trends in $\beta$ for different urban thresholds. (Inset) Historical trends in $\beta$ using a mixed threshold (50 for 1850-1930, at 200 for 1930-2020) .}
    \label{fig:si7}
\end{figure}
The key hyperparameter of the US analysis is the urban threshold: the minimum population required for a cell to be classified as urban. Setting this parameter requires balancing two competing issues across a long historical period. On the one hand, low thresholds cause large cities to merge, particularly in later years when census place populations are large. On the other hand, high thresholds exclude small cities, especially in the early years, when census place populations are small. Our threshold at 50 people is a compromise: it allows us to have small cities (of 5,000 people) across the whole time frame, while limiting mergers of large cities.

As we vary the threshold, the major difference is mechanical and concentrated in the early period. When we raise the threshold to 200, many small cities drop from the sample (Fig. \ref{fig:si7}C), leaving only a few large cities. This makes the estimated size–growth slope for that period noisy (green line, Fig. \ref{fig:si7}D). This attrition is smaller in later periods, when city populations are larger (Fig. \ref{fig:si7}C).

Apart from this truncation and its downstream noise, the results are robust to the choice of threshold. The size–growth curves have essentially the same shape across cutoffs (Fig. \ref{fig:si7}A-B). Further, the main pattern identified in the main paper --- the weakening growth advantage of large cities --- reproduces at different thresholds, with the caveat of extra noise in the early period at the 200 threshold (Fig. \ref{fig:si7}D). The findings also replicate with a time-varying threshold, which preserves the sample of small cities in the early period while reducing the mergers of large cities in the late period (Fig. \ref{fig:si7}D Inset). 
\newpage
\null
\newpage

\section{Other robustness tests}
\subsection{Choice of national borders}
\begin{figure}[h!]
\centering
\includegraphics[width=0.95\textwidth]{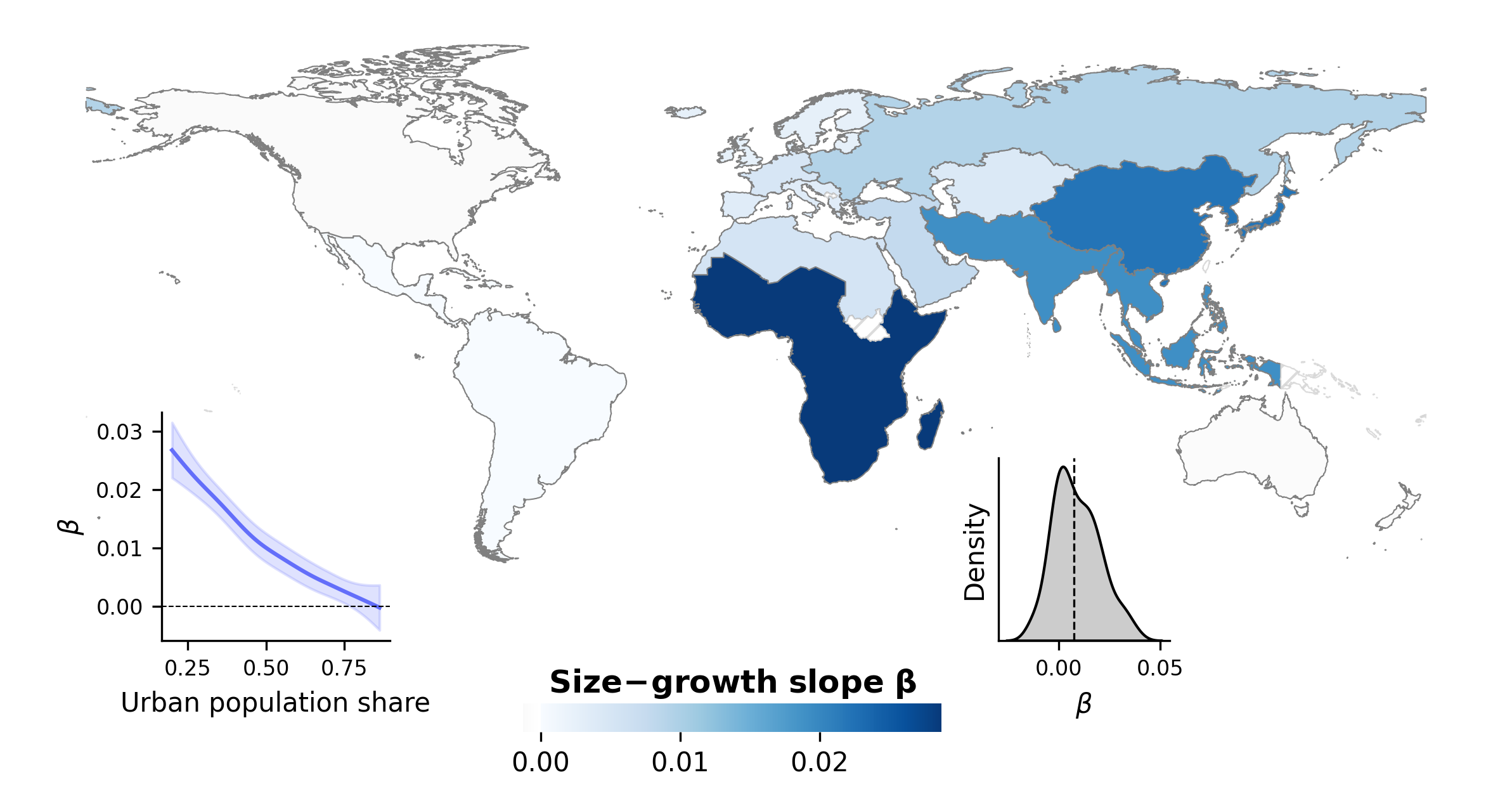}
\caption{Robustness of results to aggregation at the level of UN M49 sub-regions. The map shows the mean size-growth slope $\beta$ by sub-region, averaged over the 1975-2025 period (cf. Figure 1C in the main text). For each sub-region-year, $\beta$ is estimated by pooling cities within the sub-region and fitting the same penalized cubic B-spline used in the main text ($\lambda = 100$). Hatched areas indicate no data. The left inset shows the relationship between $\beta$ and the sub-regional urban population share, using a penalized cubic B-spline fit (cf. inset of Figure 2A in the main text). The right inset shows a kernel density estimate of the distribution of $\beta$ across sub-region-years (dashed line = median).}
\label{fig:si8}
\end{figure}

In the main paper, we define urban systems using national borders for three reasons. First, national borders constrain migration flows and infrastructure investment, which are central mechanisms of differential urban growth \cite{verbavatz2020growtha}. Second, national borders are the most commonly used delineations of urban systems in the literature on city-size distributions, making our work easier to compare with earlier studies.  Third, data and projections about population and urban population share are typically reported at the national level.

We recognize that this approximation is not exact, because some urban systems extend across borders. For this reason, we provide further evidence that the core pattern we identify---a declining growth advantage of large cities as urbanization progresses---holds qualitatively at different scales. In the main text, we show this across four broad macro-regions (Asia, Europe, Africa, and the Americas). Here, we show it across the UN M49 sub-regions.

The UN M49 classification defines 16 geographical sub-regions that are more homogeneous than continents but broader than individual countries (see the map in Fig. \ref{fig:si8}). After filtering out sub-regions with fewer than 50 cities (Polynesia and Melanesia), we are left with 14 sub-regions, yielding 126 sub-region-year observations. For each sub-region-year, we pool cities from the countries in the sub-region and estimate the size-growth slope $\beta$ with the same penalized cubic B-spline procedure used in the main text ($\lambda = 100$). The sub-regional urban population share is computed as a population-weighted average of the urban population shares of the countries within the sub-region.

Fig. \ref{fig:si8} shows that the sub-regional pattern closely resembles the country-level one. The average $\beta$ is high in Sub-Saharan Africa, Southern Asia, and Eastern Asia, while it is low in Western Europe, South America, and North America. The negative association with the urban population share remains clear: a pooled OLS regression of $\beta_{rt}$ on the sub-regional urban population share\footnote{We compute this share by taking the population-weighted average of the urban population shares of countries within the sub-region.} $u_{rt}$ yields a negative slope of $-0.0363$, with a 95\% confidence interval of $[-0.044, -0.029]$ and $R^2 = 0.436$. The coefficient is somewhat smaller in magnitude than in the pooled country-level regression ($-0.0363$ vs.\ $-0.049$), but the fit is substantially better ($R^2 = 0.436$ vs.\ $0.168$). This suggests that the relationship becomes less noisy at this intermediate spatial scale.

\subsection{Urban population share data}
In the main text, we use the urban population share data from Chen et al. \cite{chen2022updating}. A drawback of this dataset is that it relies on national statistical agencies to define ``urban'', introducing potential cross-country inconsistencies. Here, we complement the main text results with a robustness check. We replicate Table 2 from the main text, using an alternative urban population share measure from the History Database of the Global Environment (HYDE) \cite{pbl2023hyde}. HYDE is more consistent with our city definition because its urban population share estimates are derived from spatially explicit grids that combine historical demographic estimates with land-use allocation algorithms. Table \ref{tab:resp:1} shows that our main findings are robust to this alternative measure of urban population share.

\begin{table}[h!]
\centering
\begin{tabular*}{0.8\linewidth}{@{\extracolsep{\fill}}lcc}
\\[-1.8ex]\hline\hline \\[-1.8ex]Independent \textbackslash{} Dependent& \multicolumn{2}{c}{Size-growth slope} \\
\midrule
Urban population share & -0.051*** & -0.048*** \\
 & (0.004) & (0.011) \\
Country fixed effect & No & Yes \\
Observations & 891 & 891 \\
$R^2$ & 0.172 & 0.545 \\
\hline\hline \\[-1.8ex]& \multicolumn{2}{r}{$^{*}$p$<$0.1; $^{**}$p$<$0.05; $^{***}$p$<$0.01} \\
\end{tabular*}
\caption{Association between size-growth slope and urban population shares using a different dataset (\cite{pbl2023hyde}) to define the urban population shares of countries.}
\label{tab:resp:1}
\end{table}

\newpage
\section{Suburbanization}

A natural candidate mechanism to explain the declining growth advantage of large cities is suburbanization. As urbanization progresses, population growth may shift away from the urban core and toward suburban satellites, reducing the core's growth rate while increasing that of the suburbs. Since suburbs tend to be small and cores large, this reallocation of growth would flatten the size–growth curve and could therefore be an important contributor to the decline in $\beta$.

We examine this mechanism in two steps. First, we focus on the United States, comparing our geographic city definitions with functional urban areas to directly isolate the role of suburbanization. Second, we construct a simple gravity-based approximation of functional urban areas from satellite-based morphological clusters (hereafter, called ``base clusters'') and apply it globally to extend our USA-related results. 

\subsection{Definition comparison for the USA}
\begin{figure}[h!]
\centering
\includegraphics[width=0.95\textwidth]{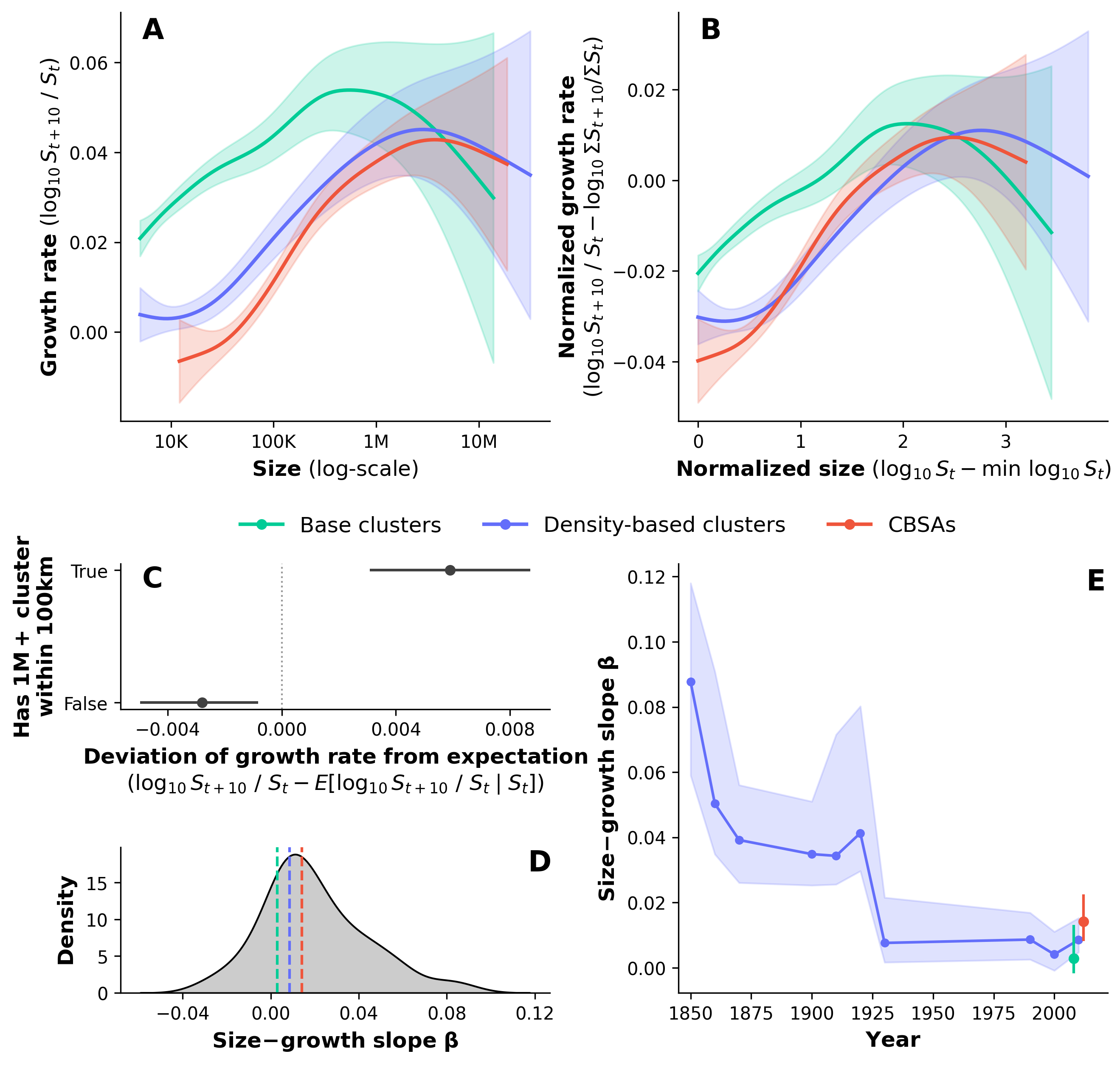}
\caption{Suburbanization as a candidate mechanism for the decline of the size-growth slope in the USA. \textbf{(A)} Size-growth curves for the USA in 2010-2020 under three city definitions: base clusters from the Global Cities Dataset, density-based clusters from the USA Cities dataset, and Core-Based Statistical Areas (CBSAs). For each definition, population growth is measured within stable decade-specific boundaries obtained with the procedure in Methods 4.2. The curves are estimated using penalized cubic B-splines (Methods 4.3.2). \textbf{(B)} The same curves after horizontal and vertical translation to compare their shapes. The $x$-axis is translated by subtracting the minimum observed log-size of each definition. The $y$-axis is translated by subtracting the average log-growth of the urban population, $\log_{10}(\sum_i S_{i,t+10}/\sum_i S_{i,t})$. \textbf{(C)} Average deviation from the expected growth among USA clusters below 1M inhabitants in 2010, split by whether they are located within 100km from a 1M$+$ cluster or not. Expected growth is the value of the size-growth curve at the cluster's size. \textbf{(D)} Distribution of country-level size-growth slopes $\beta$ in the Global Cities Dataset for 2010-2020; vertical lines indicate the USA estimates from the three definitions. \textbf{(E)} Historical evolution of the USA size-growth slope $\beta$ in the density-based data, with the 2010-2020 estimates from the base-cluster and CBSA definitions overlaid.}
\label{fig:si9}
\end{figure}

We compare three city definitions over the period 2010-2020: (i) base clusters from the Global Cities Dataset; (ii) density-based clusters from the USA Cities dataset; and (iii) Core-Based Statistical Areas (CBSAs), where the official functional definition is based on commuting ties. For all three, we estimate growth using the same boundary-matching procedure within the City Clustering Algorithm (CCA) framework described in Methods 4.2: we match 2010 clusters to their 2020 counterparts, construct a stable city boundary by taking the union of the connected components of the matching graph, and measure population change within this boundary. CBSAs were introduced in 2003, making 2010-2020 the first full census decade for which this comparison is possible. We avoid back-projecting modern functional boundaries into earlier periods because it introduces selection bias. In fact, defining a historical sample based on 2020 functional areas systematically selects for cities that grew rapidly. Small settlements that stagnated fail to meet the threshold for modern CBSA designation and are excluded, whereas those that grew rapidly meet the threshold and thus are included. Consequently, back-projecting boundaries inflates the growth rates of smaller cities.

Figure \ref{fig:si9}A shows that the density-based and CBSA curves are already fairly similar, while the base-cluster curve is visibly higher for small cities. To compare the \emph{shape} of the curves rather than their levels, we translate each curve horizontally and vertically (Fig. \ref{fig:si9}B). Specifically, for each city $i$, we plot
\begin{equation}
\tilde{s}_{it} = \log_{10} S_{it} - \min_j \log_{10} S_{jt}, \qquad
\tilde{g}_{it} = \log_{10}\left(\frac{S_{i,t+10}}{S_{it}}\right) - \log_{10}\left(\frac{\sum_j S_{j,t+10}}{\sum_j S_{jt}}\right) \ .
\end{equation}
Here, $j$ indexes cities within a given definition.
This operation aligns the lower end of the size spectrum and removes the definition-specific average growth level, but it does not rescale the curve or change its slope. After this alignment, the density-based and CBSA curves largely overlap, whereas the base-cluster curve remains higher on the left-hand side. 

This pattern is consistent with suburbanization. If small clusters near a large city grow faster than small clusters far from a large city, integrating these clusters with their larger neighbors would shift the curves in exactly this way. To test this directly, we focus on USA base clusters with fewer than 1M inhabitants. We split these clusters into two groups: those located within 100 km of a 1M$+$ cluster and those that are not. We then measure whether these small clusters grow faster or slower than expected, given their size, where the expected growth is the fitted value of the USA 2010–2020 size–growth curve at the cluster's log-size. For cluster $i$, the deviation from expectation is
\begin{equation}
\delta_i = \log_{10}\left(\frac{S_{i,t+10}}{S_{it}}\right) - \hat{f}\left(\log_{10} S_{it}\right) \ .
\end{equation}
Here, $\hat{f}$ is the fitted national size-growth curve. Positive values indicate that the cluster grew faster than typical for clusters of that size. Negative values indicate the opposite. Figure \ref{fig:si9}C shows a statistically significant positive result for the suburbanization channel: small clusters near a 1M$+$ city have positive average deviations, while small clusters far from such cities have negative average deviations. 

The remaining question is how much of the size-growth slope trends identified in the main text can be explained by this mechanism. Figure \ref{fig:si9}D-E suggests it has limited explanatory power, at least in the US. In the historical USA series, the density-based estimate of $\beta$ falls from roughly $0.09$ in 1850 to roughly $0.01$ by 2010, whereas the gap across the three 2010-2020 definitions is of the order of $0.01$ or less (Fig. \ref{fig:si9}E). Similarly, when the 2010-2020 USA estimates are placed against the cross-country distribution of $\beta$ in the Global Cities Dataset during that same decade, the spread across definitions occupies only a small part of that distribution (Fig. \ref{fig:si9}D). Thus, empirical evidence for the USA indicates that suburbanization is present and measurable, and that it pushes $\beta$ in the expected direction, but that its magnitude is small relative to other forces.

\subsection{Gravity-based approximation of functional urban areas}

\begin{figure}[h!]
\centering
\includegraphics[width=0.95\textwidth]{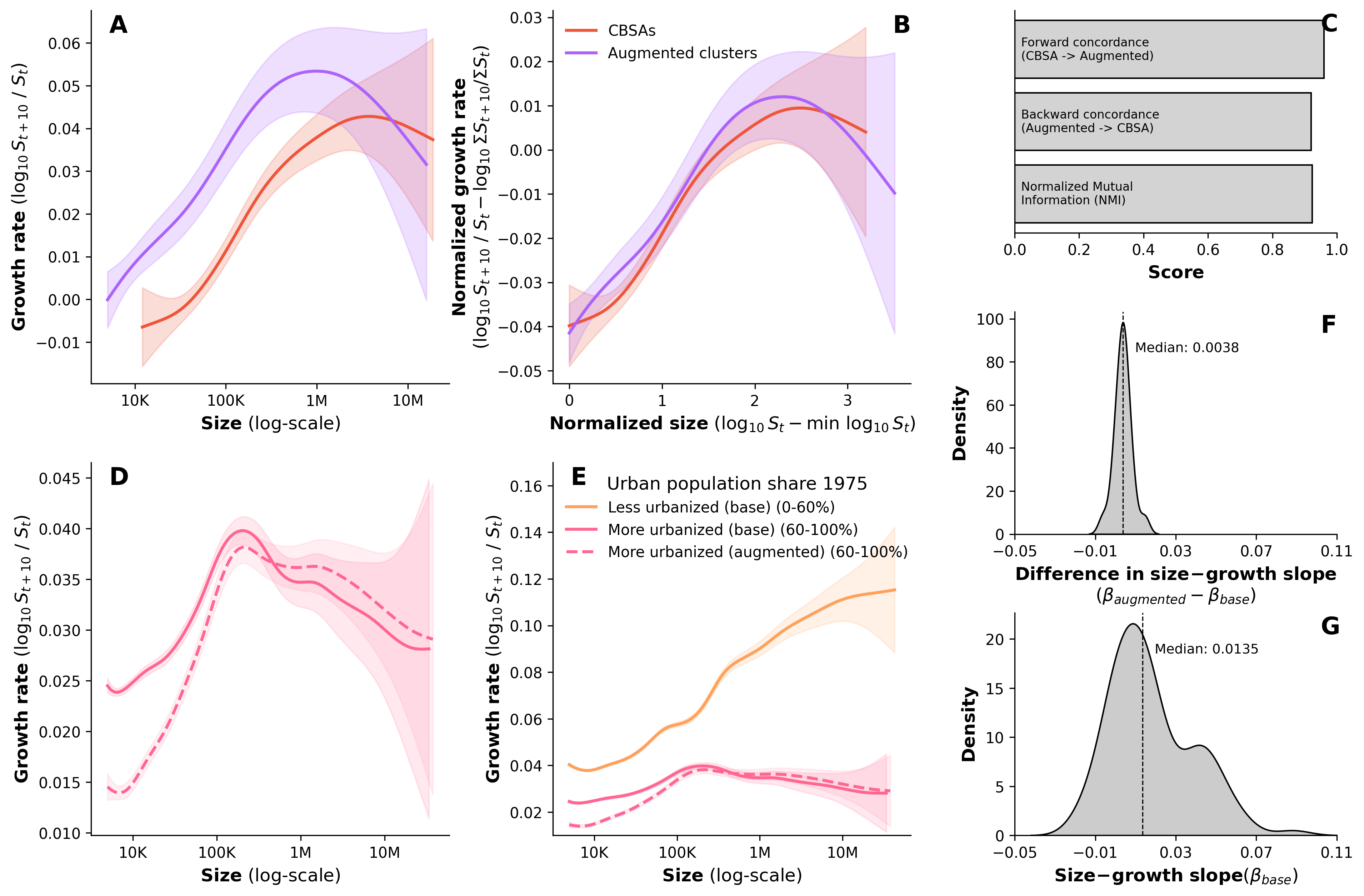}
\caption{Gravity-based approximation of functional urban areas and the magnitude of the suburbanization channel. \textbf{(A)} Size-growth curves for CBSAs and gravity-based augmented clusters in the USA 2010-2020. \textbf{(B)} Size-growth curves in (A) translated as in Fig. \ref{fig:si9}B. \textbf{(C)} Validation of gravity-based grouping against CBSAs using forward concordance, backward concordance, and normalized mutual information. \textbf{(D)} Size-growth curves for the more-urbanized group under the base-cluster and augmented definitions. Curves are estimated as in Methods 4.3.2. \textbf{(E)} Size-growth curve for the less-urbanized group under the base-cluster definition, and for the more-urbanized group under both the base-cluster and augmented definitions. \textbf{(F)} $\beta_{\text{augmented}}$ and $\beta_{\text{base}}$ denote the size-growth slopes estimated using the augmented and base-cluster definitions. The figure shows the distribution of $\beta_{\text{augmented}} - \beta_{\text{base}}$ across countries in the more-urbanized group. For comparison, we set the same $x$-range as in (G). \textbf{(G)} Distribution across all countries of $\beta_{\text{base}}$. With a mean difference of 0.0037 in (F) and a standard deviation of 0.0206 in (G), the standardized difference represents 0.18 standard deviations of the global base distribution.}
\label{fig:si10}
\end{figure}

To extend the same analysis beyond the USA, we construct a simple gravity-based approximation of functional urban areas from the base clusters. The idea is to use a gravity-based heuristic to decide whether to absorb a suburban satellite into a large metropolitan area nearby. For each ordered pair of base clusters $(c_1,c_2)$, let $d(c_1,c_2)$ denote the distance between their geometries, $S_{c}$ the population of cluster $c$, and $A(c)$ its area. We define the \emph{pull} of $c_1$ on $c_2$ and the \emph{self-pull} of a cluster $c$:
\begin{equation}
\text{pull}(c_1 \rightarrow c_2) = \frac{S_{c_1}}{d(c_1,c_2)},
\qquad
\text{self-pull}(c) = \frac{S_c}{\sqrt{A(c)}} \ .
\end{equation}
In the self-pull equation, the term $\sqrt{A(c)}$ acts as a proxy for the characteristic internal distance of the cluster. Because a settlement's radius is proportional to the square root of its area, this term captures the expected distance between random inhabitants within the city, thereby allowing for a comparison with external distances.
For each base cluster $c$, we identify the neighbor exerting the largest pull. If this largest external pull exceeds $\text{self-pull}(c)$, the base cluster is reassigned to that neighbor; otherwise, it remains separate. This procedure yields  ``augmented clusters''. Intuitively, these augmented clusters are the base clusters merged with smaller nearby base clusters, on which they exert a strong pull. Empirically, the mean distance between a merged satellite and its central cluster is 25 km (interquartile range: 5 to 40 km).

To verify the effectiveness of this gravity-based method in approximating functional urban areas, we compare the partition of the base clusters induced by the augmented clusters with that induced by CBSA membership. Specifically, we assign each base cluster $c$ two labels:
\begin{enumerate}
    \item $u(c)$ is the CBSA with which $c$ has the largest area of overlap.
    \item $v(c)$ is the augmented cluster with which $c$ has the largest area of overlap. 
\end{enumerate}
We then compute forward concordance, backward concordance, and normalized mutual information on these labels, weighting by population. Forward concordance measures the probability that two clusters that share the same CBSA label ($u(c)$) also share the same augmented cluster label ($v(c)$). Backward concordance measures the reverse: the probability that two clusters with the same augmented cluster label share the same CBSA label. Normalized mutual information (NMI) calculates the overall statistical dependence between the two partition systems. All three metrics range from 0 (completely independent partitions) to 1 (perfectly identical partitions). The resulting scores are high: forward concordance is 0.92, backward concordance is 0.96, and NMI is 0.92 (Fig. \ref{fig:si10}C), implying a close correspondence between the partitions. This correspondence is also visible in a simpler validation: the augmented clusters generate a size-growth curve that is similar in shape to the CBSA curve (Fig. \ref{fig:si10}A-B; this is particularly visible after translation).

Given this validation, we apply the same procedure to the countries in the ``more-urbanized group'' defined in the main text (60-100\% urban population share in 1975), where suburbanization is likely to be most pronounced. In Fig. \ref{fig:si10}D, we compare the size-growth curve estimated using augmented clusters to that estimated using the base clusters. As in the USA (Fig. \ref{fig:si10}B), we observe that the difference between the two curves concentrates on the left-hand side: small augmented clusters grow slower than small base clusters, consistent with suburbanization. However, this difference is small compared to the difference in size-growth curves between our more-urbanized and less-urbanized country groups (Fig. \ref{fig:si10}E). Quantitatively, the mean difference in size-growth slopes between the two definitions ($\beta_{\text{augmented}} - \beta_{\text{base}}$) represents a shift of only 0.18 standard deviations of the global distribution of base slope.

We re-estimate size-growth slopes using augmented clusters ($\beta_{\text{augmented}}$) and compare them with the original size-growth slopes estimated using base clusters ($\beta_{\text{base}}$). The country-level distribution of $\beta_{\text{augmented}} - \beta_{\text{base}}$ is shifted slightly to the right, with a median of about 0.004 (Fig. \ref{fig:si10}F). This is consistent with suburbanization making the base-cluster slopes flatter. At the same time, the flattening is small: the differences $\beta_{\text{augmented}} - \beta_{\text{base}}$ span just a fraction of the variation in $\beta_{\text{base}}$ across countries (Fig. \ref{fig:si10}G). We also re-ran the regression in Table 2, replacing $\beta_{\text{base}}$ with $\beta_{\text{augmented}}$ for countries in the more-urbanized group, and obtained similar results (Table \ref{tab:resp:2}). 

Taken together, these results show that suburbanization is a real and detectable part of the urban life cycle. However, it is quantitatively too small to explain the overall decline in the growth advantage of large cities. The life cycle documented in the main text must therefore reflect deeper forces that go beyond the spatial redistribution of population within metropolitan areas.

\begin{table}[h!]
\centering
\begin{tabular*}{0.8\linewidth}{@{\extracolsep{\fill}}lcc}
\\[-1.8ex]\hline\hline \\[-1.8ex]Independent \textbackslash{} Dependent& \multicolumn{2}{c}{Size-growth slope} \\
\midrule
Urban population share & -0.048*** & -0.056*** \\
 & (0.004) & (0.012) \\
Country fixed effect & No & Yes \\
Observations & 810 & 810 \\
$R^2$ & 0.141 & 0.531 \\
\hline\hline \\[-1.8ex]& \multicolumn{2}{r}{$^{*}$p$<$0.1; $^{**}$p$<$0.05; $^{***}$p$<$0.01} \\
\end{tabular*}
\caption{Association between country urban population share and size-growth slope $\beta$. We use augmented clusters to estimate $\beta$ for countries in the ``more-urbanized group'' (60-100\% urban population share in 1975) and the base clusters to estimate $\beta$ for the ``less-urbanized group'' (0-60\% urban population share in 1975).}
\label{tab:resp:2}
\end{table}

\bibliography{sn-bibliography}

\end{document}